 \documentclass[12pt, a4paper]{article}
\usepackage{amsmath,amssymb,amsfonts,amsthm,graphicx,placeins, float, hyperref}
\usepackage{color}
\usepackage[toc,page,title,header,titletoc]{appendix}
\usepackage{multirow,booktabs}
\usepackage[left=1in,right=1in,top=1in,bottom=1in]{geometry}

\usepackage{setspace}
\usepackage{amsmath,amssymb,amsfonts,amsthm,graphicx,placeins, float, hyperref}
\usepackage{epstopdf}  
\usepackage{color}
\usepackage{natbib}
\usepackage{hyperref}
\hypersetup{colorlinks,linkcolor={blue},citecolor={green},urlcolor={red}} 
\newcommand{\bD}{\mbox{\boldmath$D$\unboldmath}}

\newcommand{\bff}{\mbox{\boldmath$f$\unboldmath}}

\newcommand{\bI}{\mbox{\boldmath$I$\unboldmath}}

\newcommand{\bQ}{\mbox{\boldmath$Q$\unboldmath}}

\newcommand{\bs}{\mbox{\boldmath$s$\unboldmath}}

\newcommand{\by}{\mbox{\boldmath$y$\unboldmath}}
\newcommand{\bY}{\mbox{\boldmath$Y$\unboldmath}}
\newcommand{\bx}{\mbox{\boldmath$x$\unboldmath}}

\newcommand{\bz}{\mbox{\boldmath$z$\unboldmath}}

\newcommand{\bpsi}{\mbox{\boldmath$\psi$\unboldmath}}

\newcommand{\bbeta}{\mbox{\boldmath$\beta$\unboldmath}}

\newcommand{\btheta}{\mbox{\boldmath$\theta$\unboldmath}}

\newcommand{\bet}{\mbox{\boldmath$\eta$\unboldmath}}

\usepackage{amsmath}
\usepackage{stackengine}

\title{
Bayesian Semi-Parametric Spatial Dispersed Count Model for Precipitation Analysis}
\begin{document}
\maketitle

\begin{center}
\author{Mahsa Nadifar$^1$, Andriette Bekker$^{1,2} $, Mohammad Arashi$^{3,1}$, Abel Ramoelo$^2$} \\ 
\begin{flushleft}
$^1$ {{Department of Statistics}, {University of Pretoria}, {{Pretoria}, {South Africa}}}\\
$^2$ {{Centre for Environmental Studies, Department of Geography, Geoinformatics and Meteorology}, {University of Pretoria}, {{Pretoria}, {South Africa}}}\\
$^3$ {{Department of Statistics}, {Ferdowsi University of Mashhad}, {{Mashhad}, {Iran}}}\\
\end{flushleft}
\end{center}

\section*{Abstract}
The appropriateness of the Poisson model is frequently challenged when examining spatial count data marked by unbalanced distributions, over-dispersion, or under-dispersion. Moreover, traditional parametric models may inadequately capture the relationships among variables when covariates display ambiguous functional forms or when spatial patterns are intricate and indeterminate. To tackle these issues, we propose an innovative Bayesian hierarchical modeling system. This method combines non-parametric techniques with an adapted dispersed count model based on renewal theory, facilitating the effective management of unequal dispersion, non-linear correlations, and complex geographic dependencies in count data.  
We illustrate the efficacy of our strategy by applying it to lung and bronchus cancer mortality data from Iowa, emphasizing environmental and demographic factors like ozone concentrations, PM2.5, green space, and asthma prevalence. Our analysis demonstrates considerable regional heterogeneity and non-linear relationships, providing important insights into the impact of environmental and health-related factors on cancer death rates. This application highlights the significance of our methodology in public health research, where precise modeling and forecasting are essential for guiding policy and intervention efforts. 
Additionally, we performed a simulation study to assess the resilience and accuracy of the suggested method, validating its superiority in managing dispersion and capturing intricate spatial patterns relative to conventional methods. The suggested framework presents a flexible and robust instrument for geographical count analysis, offering innovative insights for academics and practitioners in disciplines such as epidemiology, environmental science, and spatial statistics.
 \\

\noindent{\bf Keywords:} Bayesian, GC distribution, semi-parametric spatial model, Thin-plate spline, SDG, lung and bronchus mortality, air pollution, precipitation. 
\begin{small}
\\
\end{small}
\noindent{\bf Mathematics Subject Classification (2010):} $\rm 62J12$, $\rm 62F15$,  $\rm 62H11$.\\

\section{Introduction}\label{Sec1}
Spatial count data play a crucial role in several fields such as disease mapping, environmental studies, ecology, sociology, criminal analysis, and public health.  While spatial generalized linear mixed models (SGLMMs; \citeauthor{baghishani2011}, \citeyear{baghishani2011}) have become popular for analyzing this type of data, there are situations when more flexibility is required due to intricate variable interactions and the presence of unknown or complex spatial patterns. 
Traditional parametric models often struggle to accurately represent nuanced characteristics of the data in these situations. Utilizing non-parametric approaches is a novel and effective strategy to address these problems. It enables us to transcend the limitations of pre-established functional frameworks and adapt to intricate interdependencies and unfamiliar interactions \citep{Illian2013}. 
By employing non-parametric approaches, we allow the data to dictate the relationships, leading to a more trustworthy approach that mitigates the risk of bias arising from erroneous assumptions. Furthermore, non-parametric models frequently demonstrate better performance than parametric models when handling complex spatial patterns
\citep{Illian2013, Gomez2020, Zhang2023}.
Therefore, it is essential to utilize adequately adaptable models. Implementing a structured additive regression (STAR) framework, as outlined by \cite{kneib2009}, enables us to represent non-linear relationships between covariates accurately and include various spatial dependencies.

The Poisson regression model, an essential component of generalized linear models, is commonly employed for analyzing count data. A significant weakness of the Poisson model is the assumption of equi-dispersion, where the conditional mean and variance are identical. In reality, numerous empirical datasets demonstrate deviations from this assumption, such as over-dispersion (variance exceeding the mean) or under-dispersion (variance less than the mean). This constraint requires the creation of more adaptable count regression models. 
In this context, the word \textit{non-equivalent dispersed counts} denotes count data that diverges from equidispersion, necessitating other modeling approaches. It is essential to address both over-dispersion and under-dispersion within a cohesive framework to appropriately represent the underlying data generating process. 
Various enhancements, such as negative binomial (NB) regression, generalized Poisson regression, and weighted Poisson models, have been formulated to address dispersion patterns \citep{Cameron2013, Ridout2004, Lord2010, Zhu2009}. 
 Generalized linear mixed models (GLMMs; \citeauthor{Breslow1993}, \citeyear{Breslow1993}) use random effects to address over-dispersion. The NB model, derived from a Poisson model with gamma-distributed random effects, is frequently utilized; nonetheless, it is inappropriate for under-dispersed counts. Hurdle models can effectively address over-dispersion and under-dispersion by integrating zero counts with positive counts through a two-part approach \citep{Cameron2013, Baetschmann2014}. Alternative methodologies encompass the weighting of the Poisson distribution \citep{Ridout2004}, the COM-Poisson distribution \citep{Lord2010}, generalized Poisson \citep{consul2006}, and the modified Poisson-Inverse Gaussian family \citep{Zhu2009}. Nonetheless, these models frequently exhibit a deficiency in interpretability or inadequately address underdispersion \citep{Zeviani2014}.

\cite{winkelmann1995} applied renewal theory to model non-equivalent dispersed counts, as described by \cite{Cox1962}. This method utilizes a non-exponential distribution for waiting times, offering enhanced flexibility via a time-varying hazard function. By adding an extra parameter, Winkelmann looked into the link between duration models and count models, which made it easier to relax the equidispersion assumption. The author noted that an increasing hazard function signifies positive duration dependency, which causes over-dispersion, whereas a decreasing hazard function indicates negative duration dependency, leading to under-dispersion. This methodology has been broadened to incorporate spatial modeling.The gamma-count (GC) model, constructed using renewal theory and incorporating gamma waiting time distribution, provides a probabilistic framework for count data while adeptly accommodating various dispersion patterns.

Recent methodological advancements have expanded the GC model to include spatial structures, hence improving its applicability to geographically referenced count data. Nadifar et al. (2023) presented a spatial extension of the GC model in continuous regions, illustrating its efficacy in capturing spatial dependency while preserving flexible dispersion attributes. Furthermore, \cite{nadifar2021pc} developed the GC structured additive regression model, incorporating penalized complexity (PC; \citeauthor{Simpson2017}, \citeyear{Simpson2017}) priors to control the dispersion parameter. Through a comprehensive sensitivity study of various prior selections, their methodology enhanced inference stability inside hierarchical Bayesian frameworks, hence expanding the usefulness of the GC model in epidemiology, environmental science, and social sciences.

The fundamental goal of this research is to create a thorough statistical framework that combines non-parametric spatial modeling with nonlinear fixed effects within a hierarchical Bayesian context to address different levels of dispersion in count data. Classical parametric count models frequently inadequately represent the intricate associations between covariates and response variables, especially in the presence of spatial dependencies and nonlinear effects. Utilizing renewal theory and structured additive regression models, we intend to develop a flexible and adaptable model that precisely captures overdispersed and underdispersed count responses, while accommodating spatial heterogeneity and nonlinear covariate influences. 
Our study improves the flexibility of count regression modeling by incorporating non-parametric spatial components and nonlinear fixed effects inside a Bayesian hierarchical framework, hence providing robust inference across diverse real-world applications. The suggested methodology enhances theoretical frameworks in count data modeling and offers practical benefits for environmental and epidemiological research.

We demonstrate the practical use of our innovation by utilizing it on two separate environmental datasets. Initially, we analyze lung and bronchus deaths to better illustrate the flexibility of our system. Lung-related diseases, frequently affected by air pollution, health determinants, and climatic factors, display regional correlations and varying degrees of dispersion in mortality statistics. We evaluate our model's capacity to reveal hidden spatial structures and nonlinear relationships within this dataset, offering an enhanced understanding of environmental health concerns.
Furthermore, we examine precipitation data from Alberta, Canada, specifically the count of days with at least 1.0 mm of precipitation in May 2024. Precipitation is essential for agriculture, water resource management, and infrastructure resilience, with its patterns displaying regional dependencies and nonlinear effects from climatic and topographic factors. Our suggested model adeptly encapsulates these intricacies, facilitating enhanced predictions and supporting decision-making in environmental planning.

The subsequent sections of this work are structured as follows: Section \ref{Sec2} offers a brief overview of the spatial GC regression model utilizing non-parametric techniques, whereas Section \ref{Sec3} delves into the methodology of the Bayesian semi-parametric spatial GC model. Section \ref{Sec4} highlights the efficacy of our proposed approach through its application to three real-world datasets. The first example involves a benchmark analysis of Mackerel egg counts, demonstrating that our model performs comparably to other established and efficient models. The second application focuses on the mortality rate of lung and bronchus cancer mortality in Iowa, USA, analyzing its relationship with environmental covariates. Lastly, we predict the number of days with precipitation exceeding the threshold of $1.0$ mm for missing locations in Alberta, Canada, showcasing the method's applicability in climatology. These examples align with fundamental Sustainable Development Goals (SDG) research, illustrating the model’s capacity to capture complex spatial dependencies and count data patterns across diverse domains.
 The performance of the proposed model is demonstrated in simulation analyses under various scenarios in Section \ref{Sec5}. Finally, we discuss the results in Section \ref{Sec6}.

\section{Semi-Parametric Spatial Flexible Count Regression Model}\label{Sec2} 
This section presents an overview of the GC distribution, which is a crucial element in the framework of semiparametric spatial flexible count regression models. Count data frequently demonstrate over-dispersion, a phenomenon that standard models such as Poisson regression fail to adequately accommodate. The gamma-distributed waiting times in the GC distribution provide a versatile option that can address both over-dispersion and under-dispersion in count data.
We will initially review the concept and characteristics of the GC distribution, clarifying its basic classical regression model for uncorrelated data. At that point, we will provide a regression model based on the GC distribution, elucidating how covariates can be related to the waiting times rather than the counts directly. Finally, we will analyze the hazard function related to gamma-distributed waiting times and its implications for modeling.
\subsection{Gamma-Count Distribution}
We start with a summary of what the GC distribution is and its key features, along with the basic regression model used for uncorrelated data.
According to \cite{Winkelmann2013}, the count and duration perspectives signify two distinct but related interpretations of the same stochastic process. The statistical distribution of cumulative waiting times uniquely defines the distribution of counts, and conversely. This relationship can be utilized to construct novel count data distributions \citep{Gonzales2011, McShane2008, Ong2015}. The Poisson distribution pertains to situations where inter-arrival times follow an exponential distribution, whereas the GC distribution is founded on gamma-distributed inter-arrival times, as suggested by \cite{winkelmann1995}.

Let there $\{v_k, k\geq 1\}$ be a sequence of waiting times that indicate the time between the $(k-1)$th and $k$th events. Given that $n=1,2,\ldots$, the arrival time of the $n$th event can be written as $\xi_n=\sum_{k=1}^n v_k$.
We define $Y_t$ as the total number of events that occur in the interval $(0,t)$. Therefore, the process $\{Y_t, t>0\}$ represents a counting process. For a fixed $t$, $Y_t$ is a count variable; the stochastic properties of this counting process, and consequently of the count variable, are entirely determined once we know the joint distribution function of the waiting times $\{v_k, k\geq 1\}$.
 Specifically, $Y_t<n$ if and only if $\xi_n>t$. Consequently, we can express this as:
 \[P(Y_t < n) = 1 - F_n(t),\]
 where $F_n(t)$ represents the cumulative distribution function of arrival time of $n$th event, $\xi_n$. Furthermore, we have
 \[P(Y_t=n) = F_n(t)-F_{n+1}(t),~~~~~~~~~~~n=0,1,\ldots .\]
 Typically, $F_n(t)$ results from convolutions of the underlying densities $v_k$, which can be analytically complex. However, employing renewal theory significantly simplifies the calculations when $v_k$ are independently and identically distributed with a standard distribution.
Assuming, $\{v_k,~k\geq 1\}$ is a sequence of independently and identically gamma-distributed random variables, $Gamma(\alpha, \beta)$, with mean $\mathrm{E}(v_k)=\alpha/\beta$ and variance $\mathrm{Var}(v_k)=\alpha/\beta^2$, we find that the number of events, $Y_t$, that occur within the interval $(0,t)$ follows a GC distribution characterized by parameters $\alpha$ and $\beta$, denoted by $Y_t\sim {\rm GC}(\alpha, \beta)$.
Here, $\alpha$ serves as the dispersion
parameter, controlling the level of dispersion in the observed counts.

 The probability mass function of $Y_t$ is given by \begin{eqnarray}\label{f2}
 P(Y_t=n)=G(n\alpha,\beta t)-G((n+1)\alpha,\beta t), ~~~~~ n=0,1,2,\ldots, 
 \end{eqnarray} 
 where
\[ G(n\alpha,\beta t)=\frac{1}{\Gamma(n\alpha)}\int_0^{\beta t}x^{n\alpha-1}e^{-x}dx,\]
and $G(n\alpha,\infty)=1$. For non-integer $\alpha$, no closed-form expression exists for $G(n\alpha,\beta t)$, complicating the computation of $P(Y_t=n)$. Notably, when $\alpha=1$, the waiting time distribution reduces to the exponential distribution, it leads to the Poisson distribution when substituting into \eqref{f2}. Thus, the GC distribution can be viewed as a flexible extension of the Poisson distribution, providing a foundational
model when setting $\alpha=1$.
The GC model provides versatility in modeling count data with differing levels of dispersion. These characteristics are obtained from the hazard function of the fundamental renewal process \citep{winkelmann1995}. Specifically, for \( 0<\alpha<1 \), the GC distribution displays over-dispersion, whereas for \( \alpha>1 \), it reveals under-dispersion. Further, the expected value of the GC distribution is expressed as  
\begin{eqnarray} \label{Egc} {\rm E}(Y_t)=\sum_{k=1}^{\infty}G(k\alpha,\beta t). \end{eqnarray}
Assuming consistent time intervals for every observation, we set \( t = 1 \) to preserve generality. We recommend \cite{Winkelmann2013} for those seeking a thorough explanation of the features of the GC distribution.  

Given that the dispersion characteristics of the GC model are intrinsically influenced by the behavior of the hazard function of interarrival times, we will present a concise summary of the hazard function for the Gamma distribution in the subsequent section.

  \subsubsection{Gamma-Distributed Waiting Times Hazard Function}  
The hazard function is essential for comprehending the dispersion characteristics of the GC model since it is directly connected to the interval time behavior that underlies the count distribution. The hazard function of the Gamma distribution specifically elucidates the dynamics of waiting times, offering insight into the underlying dependency patterns. This subsection deals with the mathematical formulation and characteristics of the hazard function for the Gamma distribution, which is essential for comprehending the mechanisms of over-dispersion and under-dispersion included in the GC model.

\begin{figure}  
    \centering  
    \includegraphics[width=0.8\linewidth]{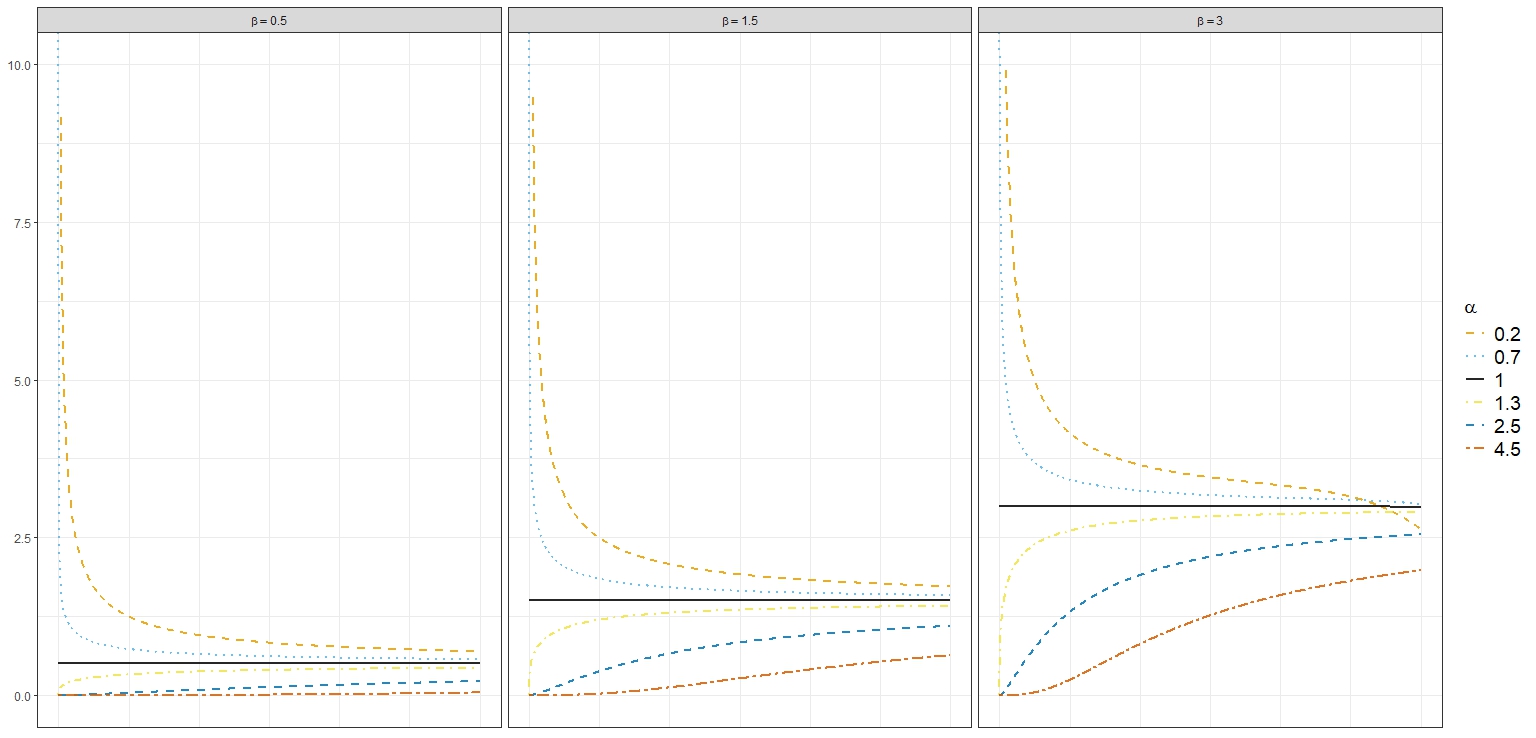}  
    \caption{Gamma distribution's hazard function with different values of shape and rate parameters.}  
    \label{hazard}  
\end{figure}

The hazard function for a Gamma distribution is represented as
\begin{eqnarray*}
\frac{1}{h_u(t)} = \int_0^\infty e^{-\gamma t}\left(1+\frac{u}{t}\right)^{\alpha-1} dt. \end{eqnarray*}  
It can be demonstrated that $h_u(t)$ is monotonically growing for $\alpha > 1$, decreasing for $\alpha < 1$, and constant for $\alpha = 1$. Figure \ref{hazard} depicts the behavior of $h_u(t)$ for diverse values of $\alpha$, emphasizing the variations in hazard trends corresponding to distinct parameter configurations.

The GC model presents an improved probability mass function for count data, enabling it to accurately represent the intricacies of waiting times and variable hazard functions. According to \cite{winkelmann1995}, a declining hazard function indicates negative duration dependency, resulting in over-dispersion in the count distribution. In contrast, a growing hazard function signifies positive duration dependence, leading to under-dispersion.
Further, figure \ref{hazard} demonstrates the impact of the rate parameter of the Gamma distribution. The rate parameter does not modify the fundamental shape of the hazard function. Still, it influences the speed of its evolution, hence regulating the rate of change without altering the underlying trends.

 \subsection{Regression Model Incorporating GC Distribution}
Let \( Y_i \) represent a count response variable coupled with a vector of covariates \( \bx_i = (1, x_{i1}, \ldots, x_{ip})' \) for \( i=1,\ldots,n \). The observations \( Y_i \) are presumed to adhere to a GC distribution, offering an adaptable structure for modeling count data with various dispersion characteristics. A primary issue in employing the GC model is the absence of a closed-form formula for its mean, as indicated in \eqref{Egc}, causing direct regression modeling based on expectation complex.  

To resolve this, we presume that the duration of the observation time is uniform for all individuals, permitting us to establish \( t=1 \) without loss of generality. Based on this premise, the GC regression model appears as follows \citep{winkelmann1995, Zeviani2014}:  
\begin{eqnarray} \label{fGCR} \log\left(\mathrm{E}\left(v_{k_i}|\bx_i\right)\right) &=& \log\left(\frac{\alpha}{\beta(\bx_i)}\right) \cr &=& - \bx'_i\bbeta = -(\beta_0 + x_{i1} \beta_1 + \ldots + x_{ip} \beta_p) = -\eta_i,~~~~~~~i=1,\ldots,n, 
\end{eqnarray}  
where \( \bbeta = (\beta_0, \beta_1, \ldots, \beta_p)' \in \mathbb{R}^{p+1} \) denotes the vector of regression coefficients. This formulation guarantees interpretability while maintaining the GC model's ability to deal with equal, over-,  and under-dispersion.  
 The indicated regression model is constructed based on the waiting times \( v_{k_i} \) instead of the observed counts \( Y_i \). This distinction stems from the essential relationship:
 \[ \mathrm{E}(Y_i|\bx_i) \propto \left(\mathrm{E}\left(v_{k_i}|\bx_i\right)\right)^{-1}, \] 
which is precisely valid when \( \alpha = 1 \). 
The negative sign in the regression equation indicates an inverse relationship between waiting times and incident counts. An increase in the expected waiting time, \( \mathrm{E}(v_{k_i}|\bx_i) \), correlates with a reduction in the expected count, \( \mathrm{E}(Y_i|\bx_i) \), illustrating how covariates influence the temporal dynamics of event occurrences. The inverse relationship is a fundamental trait of renewal-based count models, setting them apart from conventional parametric alternatives.

Given that $\{v_{k_i}, k\geq 1\}$ are independent and identically distributed random variables, we can skip the index $k$ for simplicity.
Given regressor $\bx_i$, the responses $y_i$ are conditionally independent, and their conditional mean $\mathrm{E}(Y_i|\bx_i)$ may be calculated using \eqref{Egc}. From \eqref{fGCR}, we deduce that $\beta(\bx_i)=\alpha\exp(\eta_i)$. Consequently, for a set of independent data $\{(y_i,\bx_i),~i=1,\ldots,n\}$, the GC regression model is articulated as \[y_i|\alpha,\eta_i \sim \mathrm{GC}\left(\alpha, \alpha\exp\left(\eta_i\right)\right).\]
The likelihood function appears as 
\[L(\alpha,\bet |\by)=\prod_{i=1}^{n}\left\{G\left(\alpha y_i,\alpha\exp(\eta_i)\right)\right\}-G\left(\alpha y_i + \alpha, \alpha \exp(\eta_i)\right\},\]
where $\by=(y_1,\ldots,y_n)'$ and $\bet=(\eta_1,\ldots, \eta_n)'$ represent the vectors of observed counts and linear predictors, respectively. The complexity of the likelihood function precludes straightforward forms for the maximum likelihood estimators of the parameters, hence requiring numerical optimization for parameter estimation \citep{Zeviani2014, winkelmann1995}.

\subsection{Innovation in Spatial GC Regression Model}\label{Sec2.2}
The model \eqref{fGCR} can be expanded into a semi-parametric Bayesian hierarchical framework to effectively capture spatial dependencies in count data while accommodating nonlinear interactions. Let $\bY=(Y_1, \ldots, Y_n)'$ represent the response vector, where $Y_i$ denotes the count response for the $i$th observation, which is assumed to adhere to the GC distribution. The foundational representation of waiting time, represented as $v_{k_i}$, offers a coherent framework for structured additive regression models (STAR; \citeauthor{kneib2004}, \citeyear{kneib2004}; \citeauthor{kneib2009}, \citeyear{kneib2009}). In a flexible semi-parametric regression framework, this makes it easier to include both structured and unstructured spatial elements.

The semi-parametric predictor for the generalized additive regression model is expressed as \begin{eqnarray}\label{f4}
{\rm E}(v_{k_i}|\bx_i)&=&\exp(-\eta_i), \nonumber\\ \eta_i &=& \beta_{0} + \sum_{j=1}^{p} f_j(x_{ji}) + f_s(\bs_i), \qquad i=1,\ldots,n, 
\end{eqnarray}
where $\beta_0$ denotes the intercept, $f_j(x_{ji})$ are smooth functions that encapsulate nonlinear effects of the $j$th continuous covariate for $i$th observation, and $f_s(\bs_i)$ represents the bivariate spatial effect at location $\bs_i = (s_{i1}, s_{i2})$. The spatial component $f_s(\bs_i)$ is represented using two-dimensional thin-plate splines (TPS; \citeauthor{wood2003}, \citeyear{wood2003}; \citeauthor{yue2014}, \citeyear{yue2014}), which provide considerable benefits in spatial smoothing owing to their flexibility and capacity to manage irregular spatial configurations.

We examine first-order random walk (RW1) and second-order random walk (RW2) priors for the smooth functions $f_j(x_{ji})$, which are frequently employed in Bayesian structured additive regression models (\citeauthor{rue2005}, \citeyear{rue2005}). The RW1 prior implies that adjacent function values adhere to a random walk, rendering it appropriate for identifying locally linear trends. The RW2 prior enhances this by imposing a piecewise smoothness requirement, guaranteeing that the second differences are minimal, which is especially advantageous for modeling intricate nonlinear effects. These priors offer a versatile and comprehensible framework for delineating smooth functional relationships while mitigating overfitting, rendering them particularly appropriate for structured additive regression models.

Moreover, thin-plate splines are selected as the spatial smoothing technique regarding its established potential to offer a flexible, data-driven depiction of spatial variation without necessitating explicit pre-definition of knot positions (\citeauthor{wahba1990}, \citeyear{wahba1990}). In contrast to fixed-knot splines or other basis expansions, Thin Plate Splines (TPS) optimally minimize bending energy, facilitating smooth interpolation over both regular and irregular spatial domains. TPS is especially appropriate for modeling spatial processes with unknown or complicated underlying structures, as it accommodates localized fluctuations while preventing overfitting. Moreover, TPS has been effectively utilized in Bayesian hierarchical modeling, offering a computationally efficient and statistically robust method for spatial smoothing \citep{wood2017}. TPS can be efficiently utilized for both regularly gridded and irregularly spaced spatial data, depending on the study location and data structure, rendering it a versatile option for modeling spatial interdependence in count-based outputs.

Data may be observed on regular lattices, such as $s_{11}<s_{21}<\ldots<s_{n1}~ \& ~ s_{12}<s_{22}<\ldots<s_{n2}$, where TPS might manifest during two-dimensional random walks (RW2D; \citeauthor{rue2005}, \citeyear{rue2005}). \cite{yue2010} demonstrated that TPS for regular lattice data can potentially be derived by initially specifying many second-order difference operators on lattices:
 \begin{eqnarray*} \label{IGMRF}
 && \left(f\left(s_{(i+1)1},s_{i2}\right) + f\left(s_{(i-1)1},s_{i2}\right) + f\left(s_{i1},s_{(i+1)2}\right) + \right. \cr
  &~&~~~~~ \left. 
 f\left(s_{i1},s_{(i-1)2}\right)\right) - 4f\left(s_{i1},s_{i2}\right) \cr
 &&\cr
  &~&~~~~~ = \left(\vartriangle^2_{(1,0)},\vartriangle^2_{(0,1)}\right) f\left(s_{(i-1)1},s_{(i-1)2}\right),
 \end{eqnarray*}
  where $f(\bs_{\cdot})=f(s_{1\cdot},s_{2\cdot}), \text{ and } \vartriangle_{(1,0)} \text{ and } \vartriangle_{(0,1)}$ represent the forward differences in the directions $(1,0)$ and $(0,1)$, respectively. Forward differences adhere to the multivariate Gaussian distribution as follows: \begin{eqnarray} \label{frw2d} \vartriangle^2 f(\bs_i) \sim \mathrm{N}(0, (\tau_f\bQ_f)^{-1})~~~~~~ i = 1,\ldots, n-2. \end{eqnarray}
 Furthermore, the precision matrix $\bQ_f$ may be articulated as \[\bI_n\otimes\bD'_1\bD_1+\bD'_2\bD_2\otimes\bI_n+2\bD'_1\bD_1\otimes\bD'_2\bD_2.\]
 Here, $\otimes$ denotes the Kronecker product operator, $\bI_n$ signifies a $n\times n$ identity matrix, and $\bD_1$ and $\bD_2$ are the outcomes of the decomposition of the first-order random walk (RW1) and the second-order random walk (RW2) precision matrices, respectively.

The RW2D model is inadequate for data obtained at irregularly spaced locations. The TPS estimator was obtained by \cite{yue2014} through the application of the stochastic partial differential equation (SPDE) methodology as described by \citeauthor{lindgren2011} in \citeyear{lindgren2011}. The TPS function as defined in \eqref{f4} is represented by the following expression: \begin{eqnarray*} \label{tpsirregular} \int\int\left[\left(\frac{\partial^2}{\partial s_1^2} + \frac{\partial^2}{\partial s_2^2}\right) f_s(\bs)\right]^2 ds_1 ds_2, \end{eqnarray*}
This can be derived by solving the subsequent stochastic differential equation:
\begin{eqnarray} \label{spdetps} \left(\frac{\partial^2}{\partial s_1^2} + \frac{\partial^2}{\partial s_2^2}\right) f_s(\bs) = \tau_f^{-1} \frac{d\phi(\bs)}{d\bs},
\end{eqnarray}
where $\tau_f$ represents the precision parameter and $\frac{d\phi(\bs)}{d\bs}$ denotes the spatial Gaussian white noise. 

To address \eqref{spdetps}, \cite{yue2014} utilized the SPDE methodology, implementing the finite element method (FEM) on a triangular mesh. This approach is especially efficient for managing locations that are not distributed equally, as it facilitates a more adaptable depiction of spatial fields. The outcome indicates that the TPS for the spatial effect vector $\bff_s = (f_s(\bs_1), \ldots, f_s(\bs_n))'$ adheres to a multivariate Gaussian distribution.
\begin{eqnarray} \label{tpsirregular} \bff_s | \tau_s \sim \mathcal{N}(\mathbf{0}, (\tau_s \bQ_s)^{-1}), \end{eqnarray} 
where $\tau_s$ represents the precision parameter, and $\bQ_s$ denotes the precision matrix related to a two-dimensional second-order polynomial intrinsic Gaussian Markov random field (IGMRF; \citeauthor{rue2005}, \citeyear{rue2005}). This formulation is applicable to both irregular and regular spatial domains. It is especially advantageous for irregular configurations, as it incorporates spatial smoothness and provides flexibility in modeling spatial dependencies.

The SPDE formulation offers multiple benefits compared to traditional fixed-knot methods, especially in its alignment with the data-driven characteristics of the spatial domain. This method enables accurate representation of intricate spatial relationships and differing levels of spatial continuity, which is crucial for real-world spatial datasets that may exhibit irregular patterns or missing information. The application of FEM on triangular meshes improves the model's flexibility in handling irregular spatial domains by facilitating localized adjustments in spatial smoothness. The method's adaptability renders it especially appropriate for spatial models characterized by data points that are not organized on a regular lattice.
The SPDE method demonstrates computational efficiency due to the finite element mesh, which streamlines the discretization of the spatial domain and mitigates the computational load associated with fixed-knot methods. The outcome is a versatile, effective, and statistically sound method for spatial count modeling in irregular spatial domains.

In context with the inverse relationship between gaps and the number of occurrences, the negative sign preceding $\eta$ in \eqref{f4} indicates the contrasting effects of fixed and random components on waiting times relative to their impact on counts. A longer expected time interval consistently leads to a reduction in the number of occurrences. The GC regression model is derived from essential parametric assumptions, which encompass the Poisson regression model as a specific instance under a singular parametric constraint. From \eqref{f4}, we can express the variable as $\beta(\bx_i) = \alpha\exp(\eta_i).$
The semi-parametric spatial GC (SPSGC) regression model for $i=1,\ldots, n$ observations is defined as follows:
\begin{eqnarray}
\label{GCf}
Y_i |\alpha, \beta_0,\tau_x,\tau_s &\sim&{\rm G}\left(y_i\alpha, \alpha\exp\left( \beta_{0} + \sum_{j=1}^{p} f_j(x_{ji}) + f_s(\bs_i)\right)\right)\cr
&~&~~~~~~~-{\rm G}\left(\left(y_i+1\right)\alpha, \alpha\exp\left( \beta_{0} + \sum_{j=1}^{p} f_j(x_{ji}) + f_s(\bs_i)\right)\right).
\end{eqnarray}
 In this context, $\alpha$ denotes the dispersion parameter.  

Evaluation of marginal likelihood as presented in \eqref{GCf} often involves complex integrals, which makes likelihood-based inference computationally demanding. To address this issue, the proposed models are developed within a Bayesian framework, utilizing the INLA methodology. The Bayesian GC model, while not consistently demonstrating a significant advantage over simpler alternatives such as NB regression in scenarios of over-dispersion, offers notable flexibility in addressing both over-dispersion and under-dispersion, which constitutes a clear benefit. The GC model utilizes a nested Poisson distribution to effectively capture complex patterns in waiting times, providing a robust and interpretable framework for the analysis of count data characterized by intricate underlying dependencies. This feature improves its usability across various datasets where conventional models might struggle to identify subtle patterns of dispersion.

\section{Semi-parametric Hierarchical Bayesian Learning}\label{Sec3}
 In Bayesian inference, the selection of prior distributions is crucial in influencing posterior estimates and maintaining model stability. A robust Bayesian model must incorporate prior knowledge and the inherent characteristics of the data while ensuring computational efficiency. We present a hierarchical Bayesian framework for the SPSGC regression model, incorporating structured additive components and adaptable spatial priors.

The likelihood function of the SPSGC model in \eqref{GCf} encompasses a comprehensive array of parameters, hyperparameters, and latent variables, which include the dispersion parameter $\alpha$, the fixed effect $\beta_0$, the spatial effect $\bff_s(s_1, s_2)$, the non-linear covariate effects $\bff_{x}(\bx) = \left(f_x(\bx_1), \ldots, f_x(\bx_K)\right)'$, along with the precision parameters $\tau_s$ and $\tau_x$. Choosing suitable prior distributions for these components is essential, as it directly affects posterior inference and model interpretability.

To implement regularization while maintaining flexibility, we utilize the penalized complexity (PC) prior for the dispersion parameter $\alpha$, as outlined in \cite{nadifar2021pc} as
\begin{eqnarray}
    \label{pcalpha} 
\pi(\alpha)&=&\lambda e^{-\lambda \sqrt{-2\log\Gamma(\alpha)+2(\alpha-1)\psi(\alpha)}}\mid\frac{\left(\alpha-1\right)\frac{d}{d\alpha}\psi\left(\alpha\right)}{\sqrt{-2\log\Gamma(\alpha)+2(\alpha-1)\psi(\alpha)}}\mid,
\end{eqnarray}
where $\psi(\cdot)$ is the digamma function \citep{Abramowitz1988}. Their outcomes demonstrate that the PC prior adeptly balances complexity and simplicity, surpassing alternative priors in managing overdispersion and underdispersion. PC priors \citep{Simpson2017} are explicitly formulated to punish divergences from a simpler baseline model while permitting adequate flexibility to account for intrinsic variability in the data.

To represent the latent fixed effects, $\bff_x(\cdot)$, we utilize RW1 and RW2, which enforce smoothness requirements on the functional estimates. The RW1 prior presupposes a local Markov property characterized by a first-order difference structure:
\begin{equation}\label{rw1}
x_i | x_{i-1} \sim \mathcal{N}(x_{i-1}, \tau^{-1}), \end{equation}
 where $\tau$ regulates the smoothness of the process. 
The RW2 prior is established to impose a more stringent smoothness constraint, utilizing second-order differences.
\begin{equation}\label{rw2}
x_i | x_{i-1}, x_{i-2} \sim \mathcal{N}(2x_{i-1} - x_{i-2}, \tau^{-1}), \end{equation}
thereby imposing a penalty on departures from a linear trajectory. Moreover, we consider Gaussian prior with zero mean and variance 100 for linear fixed effects.

In contrast to \cite{nadifar2021pc, nadifar2023}, we utilize thin-plate splines (TPS) to model the spatial effect $\bff_s(s_1, s_2)$, as detailed in Section \ref{Sec2.2}. TPS priors offer a versatile non-parametric method for modeling intricate spatial connections without enforcing strict structural constraints. This decision is driven by the necessity to accurately depict spatial variety while facilitating seamless transitions across spatial areas.

The precision parameters $\tau_s$ and $\tau_x$, which regulate spatial and non-linear effects, are allocated PC priors specified in \citep{Simpson2017}. This decision is justified by two reasons: (i) it adheres to the notion of penalizing needless complexity while maintaining fundamental spatial structures, and (ii) it reduces overfitting by discouraging excessive variability in functional components. Although different informative priors may be derived from expert insight, their impact on posterior inference requires sensitivity analysis to evaluate consistency.

Regarding the likelihood function specified in \eqref{GCf} and these prior distributions in \eqref{pcalpha}, \eqref{frw2d}, \eqref{tpsirregular}, \eqref{rw1}, \eqref{rw2}, and PC precision prior \citep{Simpson2017}, we establish a coherent and interpretable Bayesian framework that integrates structured prior knowledge with data-driven inference. Under the assumption of prior independence among parameters, the joint posterior distribution of the proposed model is expressed as:
\begin{eqnarray} 
\label{bayesianGCS+} 
\pi(\alpha, \bff_{x}, \bff_s, \tau_{s}, \tau_{x}|\by)&\propto &\prod_{i=1}^{n} \left\{
{\rm G}\left(y_i\alpha, \alpha\exp\left(\beta_{0} + \sum_{j=1}^{p} f_j(x_{ji}) + f_s(\bs_i)\right)\right)\right.\cr
 && -\left.{\rm G}\left(\left(y_i+1\right)\alpha, \alpha\exp\left(\beta_{0} + \sum_{j=1}^{p} f_j(x_{ji}) + f_s(\bs_i)\right)\right)\right\}\cr 
 &&~~~~~~~~~~~~~
\pi(\alpha)\prod_{k=1}^{K}\pi_{{\rm RW}}(\bff_{z}(\bz_k))\pi(\tau_{z})\pi_{{\rm TPS}}(\bff_s)\pi(\tau_{s}). 
\end{eqnarray}
This hierarchical Bayesian framework improves interpretability and offers a strong method for capturing spatial and non-linear correlation count data with different levels of dispersion, enabling reliable and efficient inference.
Traditionally, inference for models \eqref{bayesianGCS+} is accomplished using Markov Chain Monte Carlo (MCMC) sampling. However, it is well-known that MCMC methods encounter issues related to both convergence and computational time when applied to such complex models \citep{rue2009}. In particular, implementing Bayesian inference through MCMC algorithms for large spatial data could take several hours or even days to complete. To address this challenge, \cite{rue2009} introduced the Integrated Nested Laplace Approximation (INLA) method, a deterministic algorithm capable of providing accurate results in seconds or minutes. INLA combines Laplace approximations and numerical integration efficiently to approximate posterior marginal distributions. Let $\bpsi$ denote the $\ell\times 1$ vector of all the latent Gaussian variables, which is $(\bbeta, \bff_{x}, \bff_s)^\prime$, with $\ell$ determined by the specific model. Furthermore, let $\btheta$ represent the vector of hyper-parameters, which is $(\alpha, \tau_z, \tau_s)^\prime$ for model \eqref{bayesianGCS+}.  In practice, our primary interest lies in the marginal posterior distributions for the elements of the latent Gaussian variables and hyper-parameters, given by:
\[
\pi(\psi_j|\by)=\int\pi(\psi_j,\btheta|\by)d\btheta=\int\pi(\psi_j|\btheta,\by)\pi(\btheta|\by)d\btheta,~~j=1,\ldots,\ell,
\]
and
\[
\pi(\theta_k|\by)=\int\pi(\btheta|\by)d\btheta_{-k},~~k=1,2,3,
\]
where $\btheta_{-k}$ represents $\btheta$ with the $k$th element removed. The essential feature of INLA is to use this form to construct nested approximations 
\begin{eqnarray*} 
\tilde{\pi}(\psi_j|\by)&=&\int\tilde{\pi}(\psi_j|\btheta,\by)\tilde{\pi}(\btheta|\by)d\btheta, \\ 
\tilde{\pi}(\theta_k|\by)&=&\int\tilde{\pi}(\btheta|\by)d\btheta_{-k}, 
\end{eqnarray*} 
where Laplace approximation is applied to carry out the integrations required for evaluation of $\tilde{\pi}(\psi_j|\btheta,\by)$. 
A crucial success of INLA is its ability to compute model comparison criteria, such as deviance information criterion (DIC; \citeauthor{Spiegelhalter2002}, \citeyear{Spiegelhalter2002}) and Watanabe-Akaike information criterion (WAIC; \citeauthor{Watanabe2012}, \citeyear{Watanabe2012}; \citeauthor{Gelman2013}, \citeyear{Gelman2013}), as well as various predictive measures, e.g., conditional predictive ordinate (CPO; \citeauthor{Pettit1990}, \citeyear{Pettit1990}) and Log-score (LS; \citeauthor{Adrion2012}, \citeyear{Adrion2012}) to compare the complexity and fit of different possible models. Our proposed GC model has already been implemented in the \texttt{R-INLA} package as a \texttt{family} argument with the name "\texttt{gammacount}". 

\section{Data Evaluation}\label{Sec4}
 We evaluate the efficacy of the proposed flexible semi-parametric spatial count model using three separate case studies to assess its practical applicability. The initial example functions as a benchmark instance, extensively utilized in non-parametric spatial analysis \citep{wood2017, yue2014}, offering a standard reference for model assessment.  
The second application uses a novel data set on lung and bronchus deaths, which incorporates environmental factors, to examine geographical correlations in mortality statistics in several areas. This dataset demonstrates the model's capacity to manage real-world health data, including diverse environmental elements that affect mortality. 
The final application employs a novel dataset regarding the frequency of days with a minimum of $1.0$ mm of precipitation in Alberta, Canada, throughout May 2024. This dataset, examined for the first time, enables a predictive evaluation of absent observations in areas that satisfy this precipitation requirement. These forecasts are essential for comprehending precipitation patterns and their ramifications in climate change research.

These case studies demonstrate the model's relevance to regular and irregular spatial domains and its efficacy in comparison to conventional count models. To evaluate the merits of the new technique, we also examine established alternatives, including the Poisson (Pois), generalized Poisson (GP; \citeauthor{consul2006}, \citeyear{consul2006}), and negative binomial (NB) models.

\subsection{Benchmark Data Analysis}
Excessive exploitation of commercially viable fish populations has resulted in a significant problem in implementing efficient management strategies. To tackle this issue, evaluating the quantity of eggs generated by a population and the number of mature fish needed to create this quantity is essential. The example provided in this section pertains to data obtained from a mackerel egg survey conducted in 1992.
This dataset is available in the \texttt{R} package \texttt{gamair}.

\begin{figure}[h!]
    \centering
\begin{tabular}{cc}
      \includegraphics[width=0.45\textwidth,height=7cm]{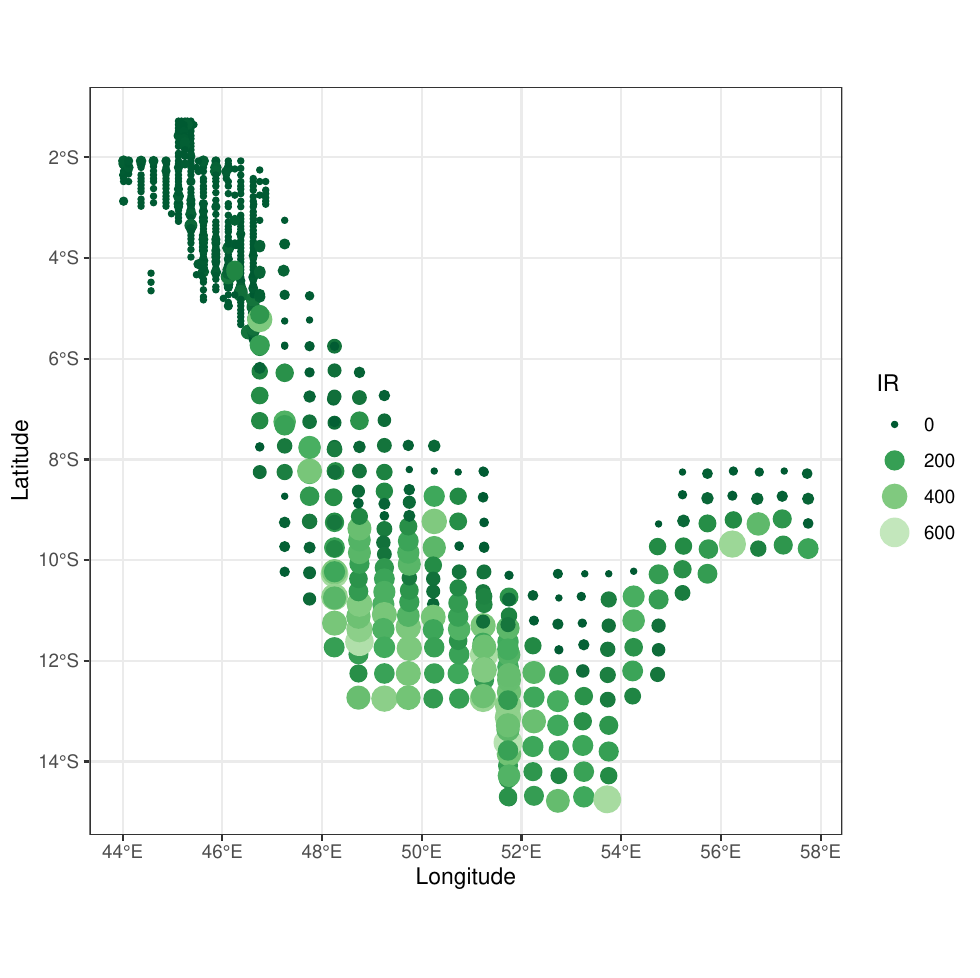} &
    \includegraphics[width=0.45\textwidth,height=7cm]{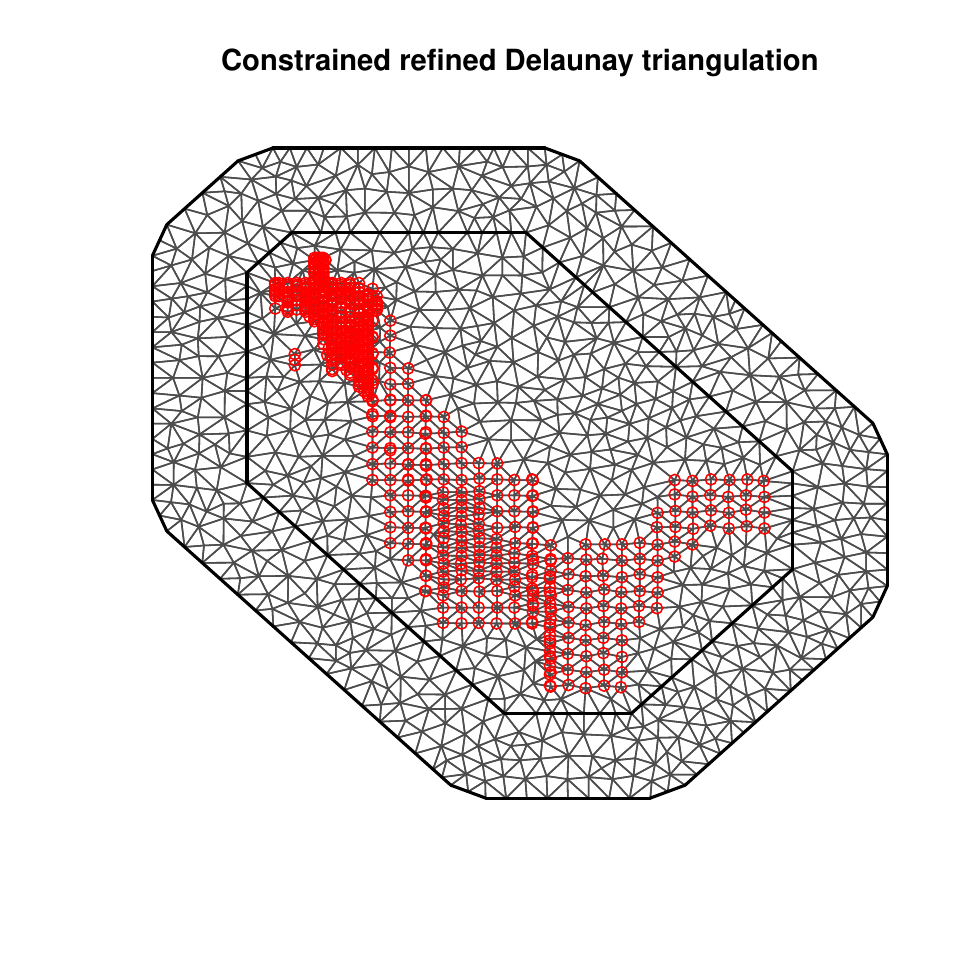}  
\end{tabular}
    \caption{Mackerel egg survey: location of the sampling and the corresponding egg counts of mackerel eggs on the sea surface in 1992  (left), and triangular mesh built for TPS with the location of the sampling (right).}
    \label{fig:data1}
\end{figure}

The goal variable in the present instance is the abundance of mackerel eggs in the net at each sample point. Our goal is to find out how the response variable, the number of eggs, is related to the fixed effects, which are salinity ($x_1$), distance from the 200-meter seabed contour ($x_2$), and water temperature at a depth of 20 meters ($z$). Figure \ref{fig:data1} exhibits the location of the sampling and the corresponding egg counts of mackerel eggs.
For the investigation, we are examining the following semi-parametric predictor:
\begin{eqnarray*}
    \eta_i = \log({\rm E}_i) + \beta_1x_{1i} + \beta_2x_{2i} + f_z(z_i) + f_s(s_{1i}, s_{2i}),~~~~~~ i = 1, \ldots,634, \end{eqnarray*} 
as an offset, ${\rm E}_i$ is the area of each net; $\beta_1$ and $\beta_2$ are the linear effects of salinity and distance from the 200-meter seabed contour, respectively; $f_z(\cdot)$ is the nonlinear effect of water temperature at a depth of 20 meters, and $f_s(\cdot)$ is the spatial effect at a point with longitudinal $s_1$ and latitudinal $s_2$.

\subsubsection{Results of Mackerel egg survey}
The model selection criteria, which include DIC and WAIC, are described in Table \ref{TabdicME}. The WAIC indicates that NB performs better than the rest, despite the DIC identifying GC as the top performer. Consequently, although our main emphasis is on the GC and NB models for estimation, we present outcomes for all four models to ensure thorough comparison and assessment.
\begin{table}[h!]
\centering\caption{\label{TabdicME}Mackerel egg survey: computed values of DIC and WAIC criteria for count models.}
\begin{tabular}{ccc}
  \hline
Model &  DIC & WAIC  \\ 
  \hline
GC &  \textbf{3051.76} & 2631.04  \\ 
  Pois &  3230.28 & 2841.69 \\ 
  NB &  3119.96 &\textbf{ 2608.42}  \\ 
  GP & 3117.44  &  2634.95 \\ 
   \hline
\end{tabular}
\end{table}

\begin{figure}[h!]
    \centering
    \includegraphics[width=0.9\textwidth,height=7cm]{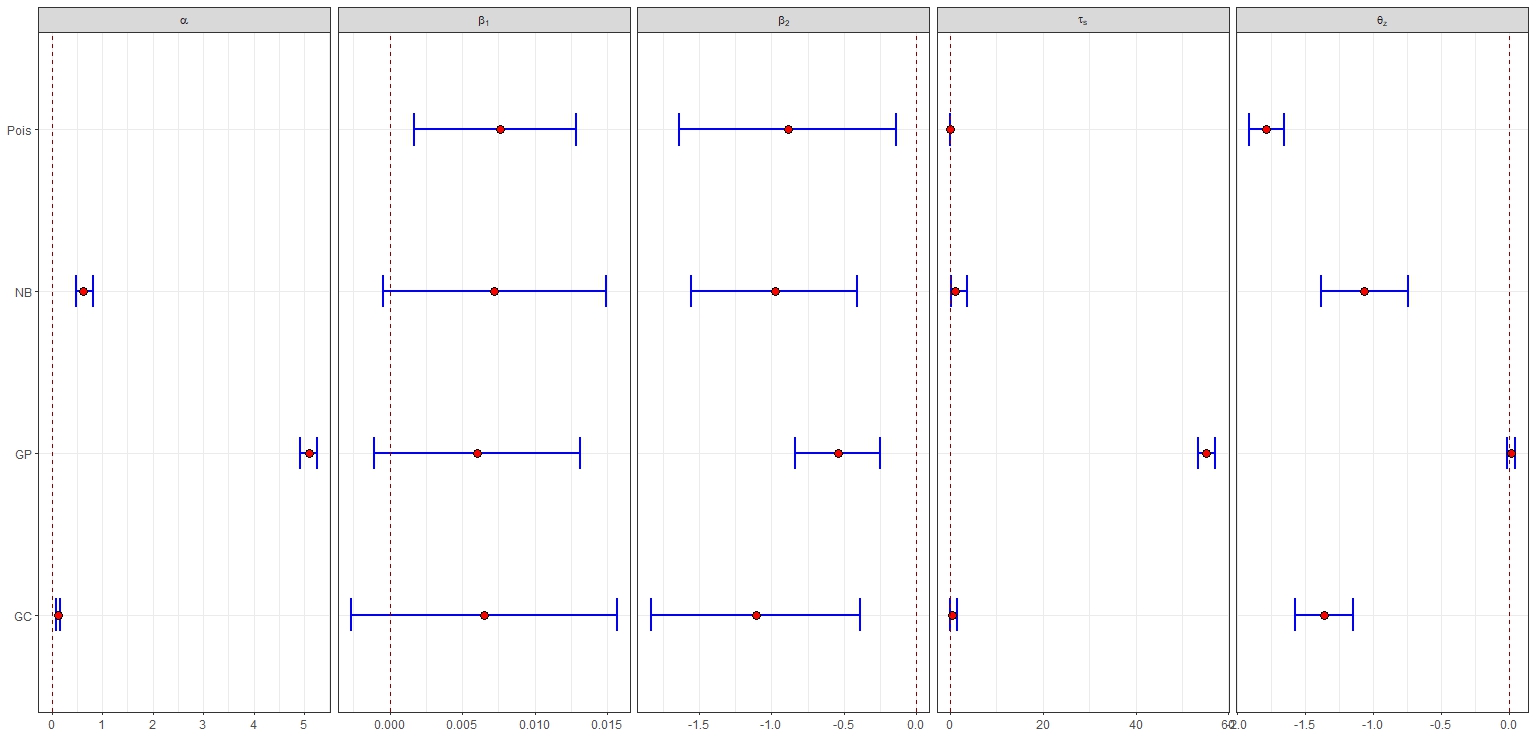}
    \caption{Mackerel egg survey: posterior mean estimates (point) and 95\% credible intervals (lines) of parameters and hyper-parameters for different count models.}
    \label{figparamhatME}
\end{figure}

The posterior mean estimates (points) and 95\% credible intervals (lines) for fixed effects, accuracy, and dispersion parameters for the different count models are displayed in Figure \ref{figparamhatME}. The estimated dispersion parameter for GC and the over-dispersion parameter for NB are shown in the leftmost column. For the GC model, the estimated amount $\alpha$ is significantly less than 0.2, indicating over-dispersion, which is consistent with the outcome shown in Table \ref{TabdicME}. The impact of salinity is considered negligible according to the chosen models, GC and NB. On the other hand, the distance estimate from the 200-meter seabed contour is similar for all models, showing the opposite impact, with NB adding greater uncertainty. The GP model displays comparable behavior to other models in estimating fixed effects, although it reveals more uncertainty regarding its precision parameter of the spatial impact.

\begin{figure}[h!]
    \centering   \includegraphics[width=0.9\textwidth,height=7cm]{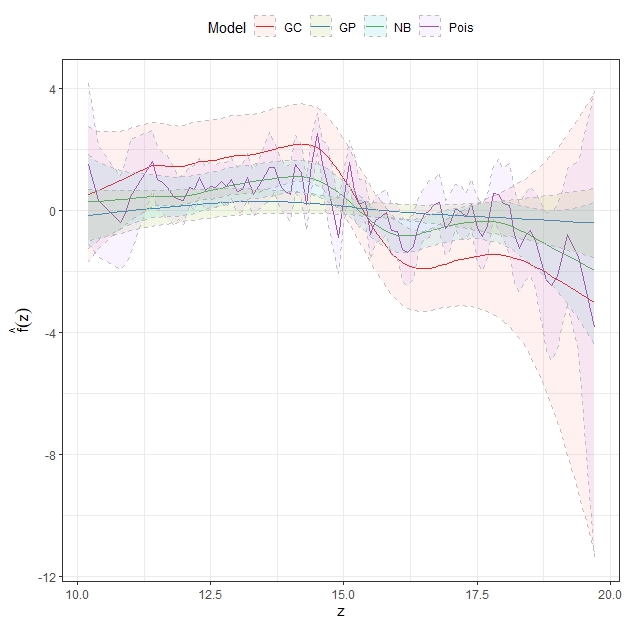}  
    \caption{ Mackerel egg survey:  Fitted curve for a nonlinear effect of temperature.}

    \label{fig:fz}
\end{figure}

The fitted curve for the nonlinear effect of water temperature at a depth of 20 meters is shown in Figure \ref{fig:fz}, which indicates that GC and NB perform similarly in extracting this nonlinear effect. However, the behavior of the Poisson model in retrieving the effect of water temperature is rough. Furthermore, the GP model suggests that temperature influence may diminish to zero, so distinguishing it from the other models. 

\begin{figure}[h!]
    \centering
\begin{tabular}{cc}
      \includegraphics[width=0.45\textwidth,height=7cm]{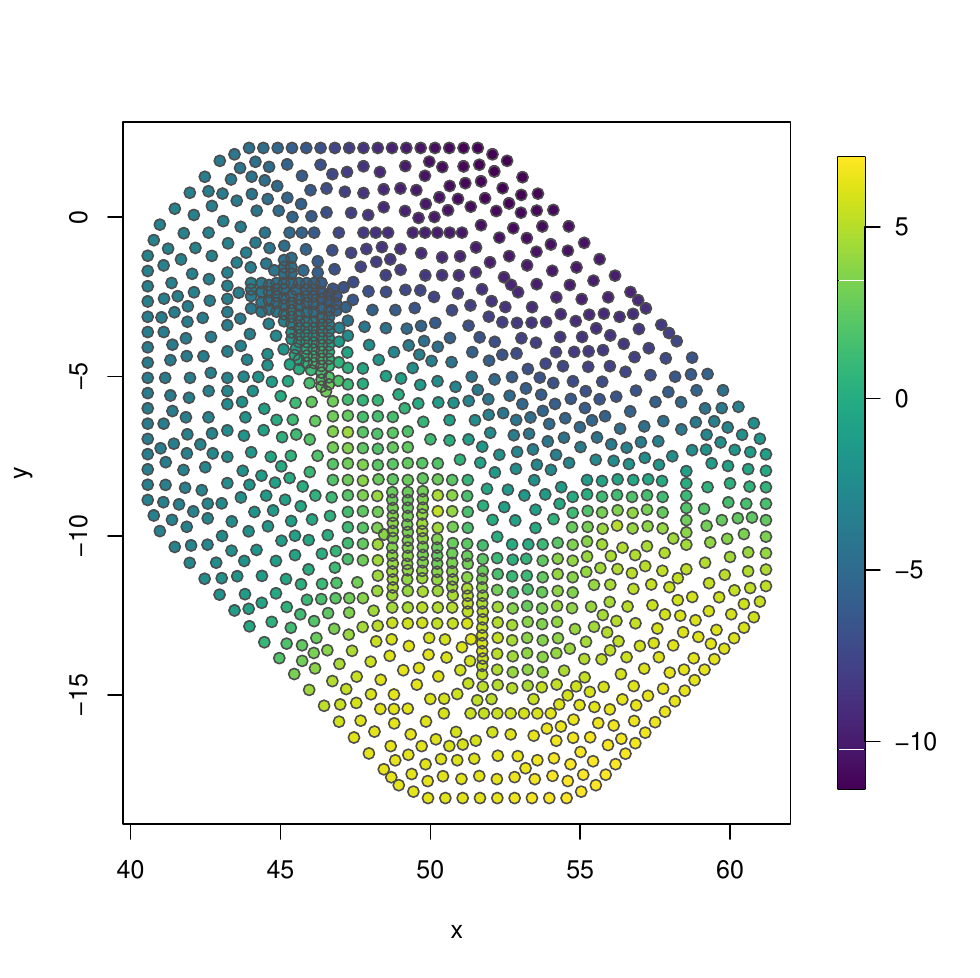} &
    \includegraphics[width=0.45\textwidth,height=7cm]{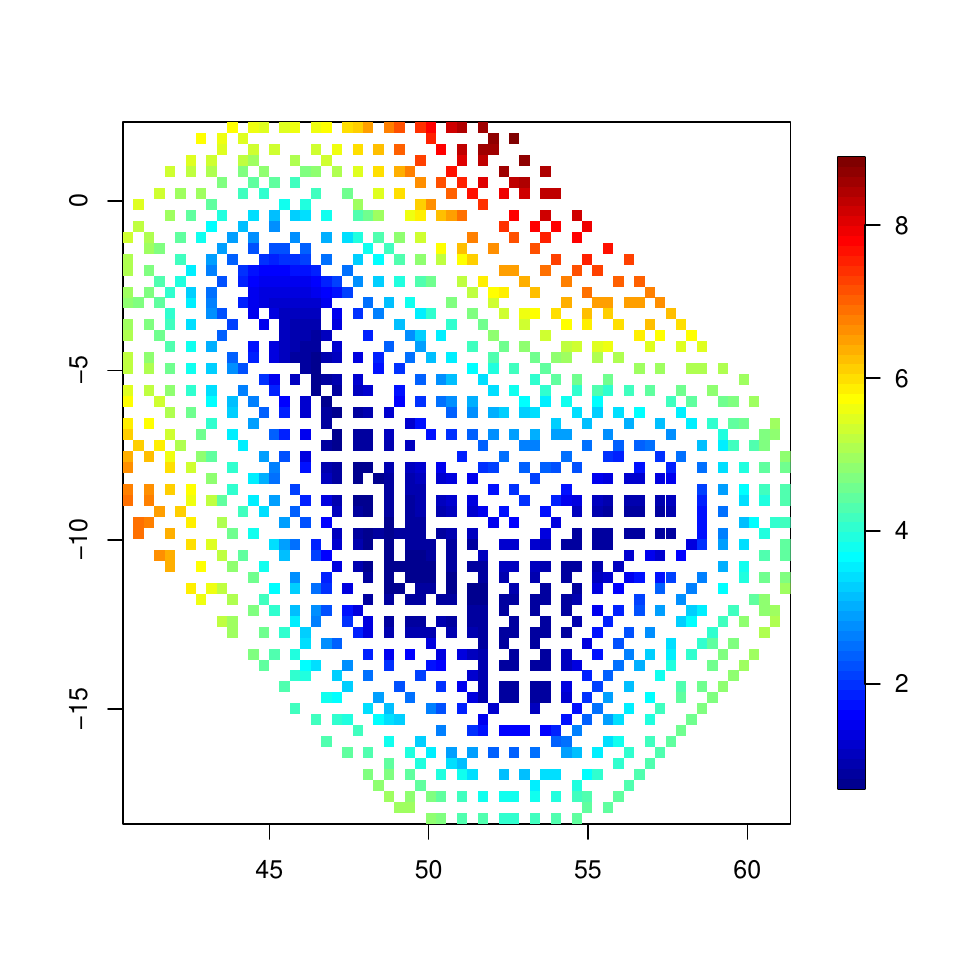}\\  
          \includegraphics[width=0.45\textwidth,height=7cm]{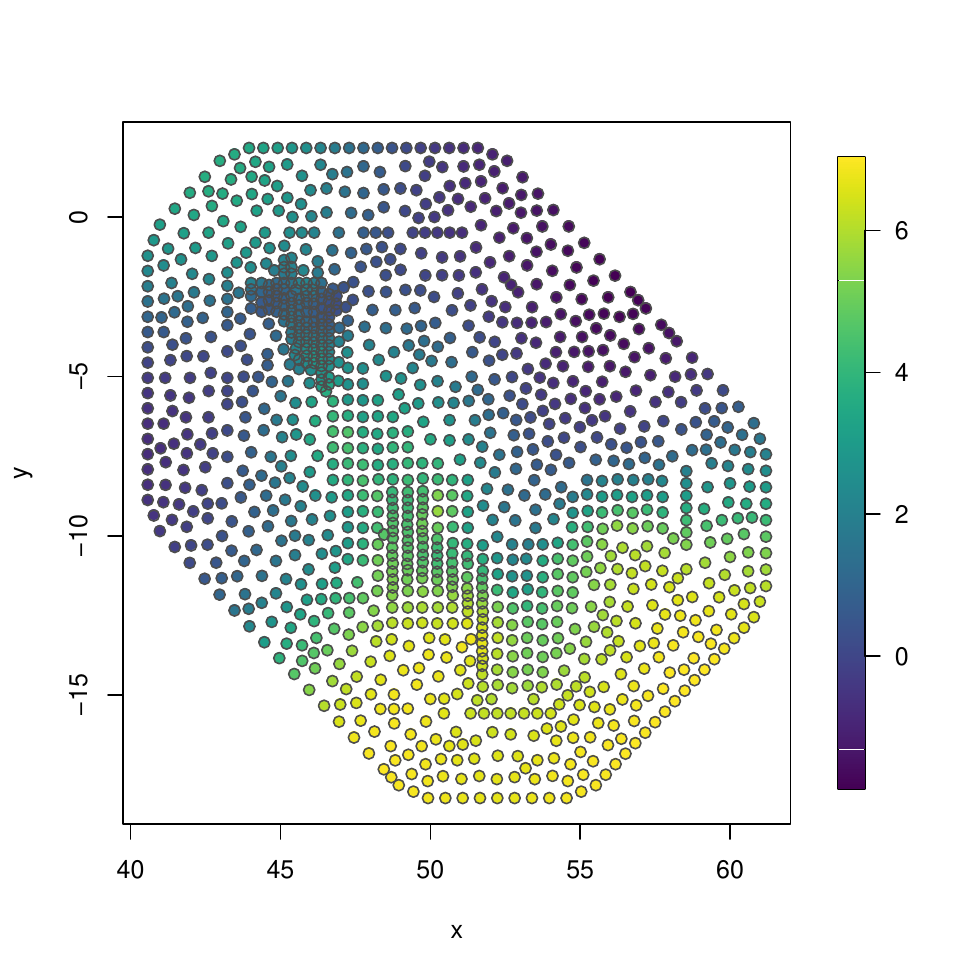} &
    \includegraphics[width=0.45\textwidth,height=7cm]{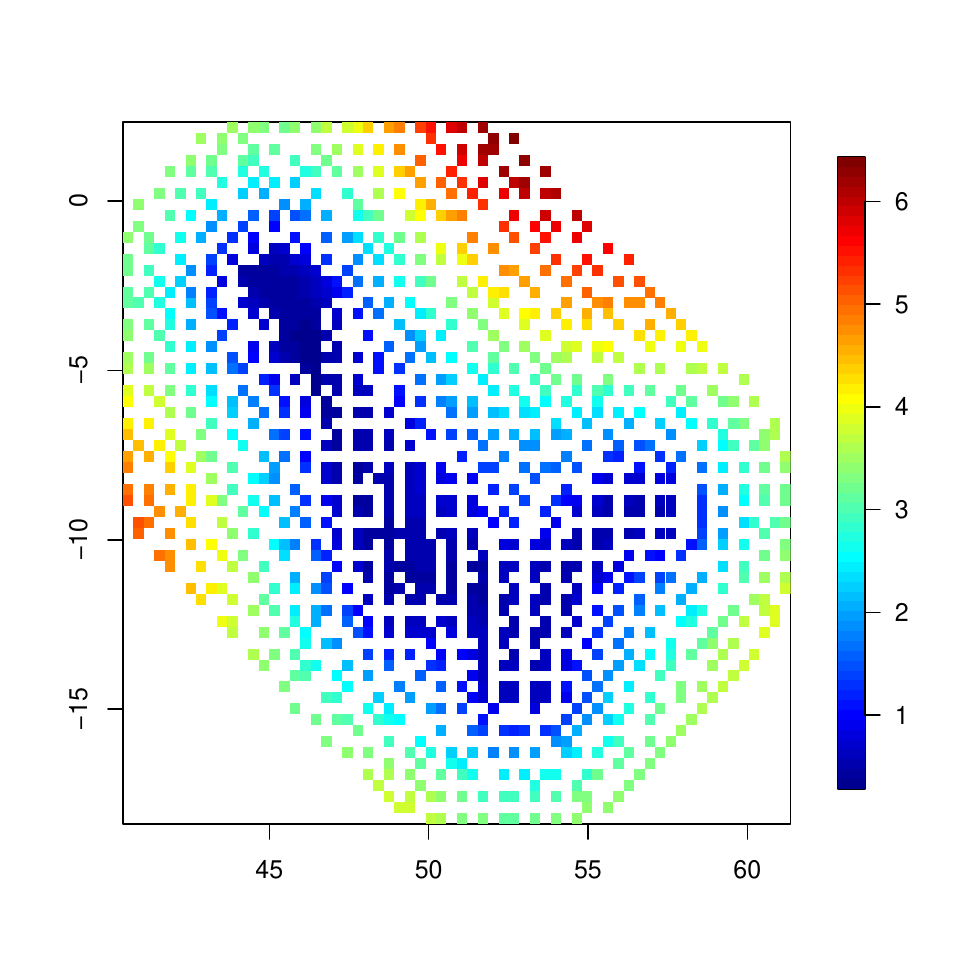}
\end{tabular}
\caption{ Mackerel egg survey: first row: posterior mean of the spatial effect (left) and its standard deviation (right) of the GC model; second row: posterior mean (left) and standard deviation (right) of the NB model.}

    \label{fig:fs}
\end{figure}

The posterior mean and standard deviation of the spatial effect are shown in Figure \ref{fig:fs}. As seen, while uncertainty is almost equal throughout all regions, the spatial dependency is more noticeable in southern regions than in northern parts.
 \subsection{Lung and Bronchus Mortality}
This application examines lung and bronchus cancer mortality in Iowa from 2018 to 2022, utilizing publicly accessible data from the U.S. Cancer Statistics database. The dataset, available at \url{https://gis.cdc.gov/Cancer}, includes the total fatalities resulting from lung and bronchus cancer ($y$) throughout 99 districts in Iowa.

The analysis includes five principal covariates: the logarithm of the population size for each district (\texttt{log(Pop)}), and four environmental and health-related factors: the percentile rank of the annual mean days exceeding the ozone regulatory standard (\texttt{O3}), the percentile rank of the annual mean days exceeding the PM2.5 regulatory standard (\texttt{PM2.5}), the percentile rank of the proportion of each district’s area within a 1-mile radius of green space (\texttt{Park}), and the percentile rank of individuals diagnosed with asthma (\texttt{Asthma}).

Ozone and PM2.5 are essential environmental indicators of air pollution, closely associated with lung cancer risk. Likewise, \texttt{Park} indicates the possible protective benefits of exposure to green space, while \texttt{Asthma} acts as a replacement for pre-existing respiratory diseases that could increase susceptibility. Data for these covariates were obtained from the Registry of the Agency for Toxic Substances and Disease Registry (\url{https://www.atsdr.cdc.gov/placeandhealth/svi}), highlighting their significance as environmental factors related to health.

This study seeks to examine the spatial variation in cancer mortality by incorporating demographic and environmental variables.
This study examines the impact of population density and air quality, both of which are recognized as key factors affecting cancer incidence. This investigation offers essential insights into the relationship between environmental risk variables and public health outcomes in Iowa.
\begin{figure}
    \centering
\begin{tabular}{cc}
    \includegraphics[width=0.55\linewidth]{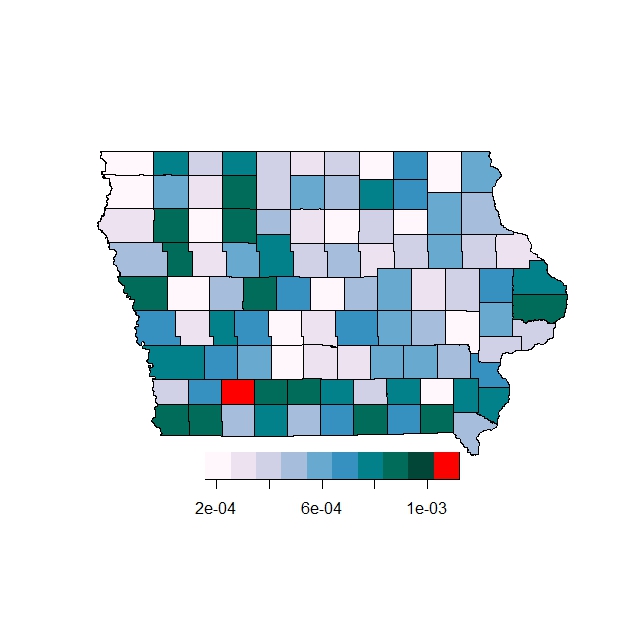}
     &     \includegraphics[width=0.4\linewidth]{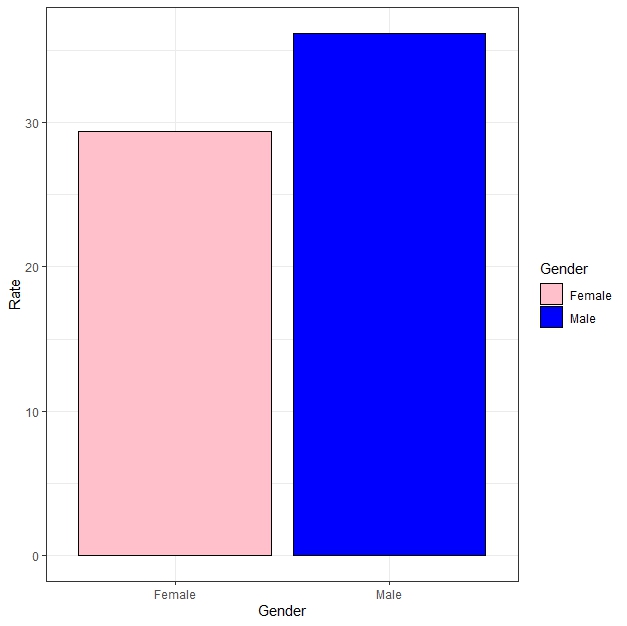}
 \\ 
\end{tabular}
    \caption{The mortality rate of lung and bronchus cancer in Iowa based on the region (left), and gender (right).}
    \label{fig:iowa}
\end{figure}
Figure \ref{fig:iowa} depicts a map of mortality rates for lung and bronchus cancer across the districts of Iowa. The mortality rate is determined by dividing the number of reported deaths by the total population of the corresponding district and multiplying by 100 ($\mathrm{IR}=\frac{y_i}{population_i}\times 100$). The picture presents a detailed analysis of cancer mortality rates by gender, distinguishing both male and female rates. The findings indicate an elevated death rate in the southern and southwestern regions of Iowa, with males consistently demonstrating greater mortality rates than females. Significantly, data regarding cancer mortality is absent for Adams County, indicated in red on the map. In the following analysis, we forecast the absent mortality rate for this county.

The observed cases are presumed to fit the $\mathrm{SPSGC}$ model conditional on the $i$-th district effect, with $\eta_i$ provided by  
\[\eta_i=\beta_0+\log({\rm Pop}_i)\beta_{{\rm Pop}}+f_{{\rm O3}}({\rm O3}_i)+f_{{\rm PM2.5}}({\rm PM2.5}_i)+f_{{\rm Park}}({\rm Park}_i)+f_{{\rm Asthma}}({\rm Asthma}_i)+f_S(\bs_i)+\epsilon_{s_i},~~~i=1,\ldots , 99,\]
where $f_{{\rm O3}}(\cdot)$, $f_{{\rm PM2.5}}(\cdot)$, $f_{{\rm Park}}(\cdot)$, $f_{{\rm Asthma}}(\cdot)$, and $f_S(\cdot)$ are smoothing functional effects of the environmental and health covariates ${\rm O3}$, ${\rm PM2.5}$, ${\rm Park}$, ${\rm Asthma}$, and a RW2D spatial latent variable, respectively. These consequences are described in full in Section \ref{Sec2.2}.
\subsubsection{Results of Lung and Bronchus Mortality }
DIC and WAIC criteria values for proposed models with different chosen pairings of priors are shown in Table \ref{Tab:diclung}.

\begin{table}[h!]
\centering\caption{\label{Tab:diclung}Lung and Bronchus Mortality: computed values of DIC and WAIC criteria for suggested models.}
\begin{tabular}{ccc}
  \hline
Model &  DIC & WAIC  \\ 
  \hline
GC &  \textbf{720.339} &\textbf{ 712.212}  \\ 
  Pois &  723.846 & 716.031 \\ 
  NB &  779.160 & 783.239  \\ 
  GP & 880.731  &  1798.716 \\ 
   \hline
\end{tabular}
\end{table}
Comparing the GC model to other models like Poisson, generalized Poisson, and NB, the former shows better accuracy.
The GC model performs noticeably better than the NB and generalized Poisson models, even though it occasionally performs similarly to the Poisson model. These results show how important it is to properly account for dispersion in count data, as shown by the GC model's consistently high performance in a range of situations.
 
\begin{figure}
    \centering
    \includegraphics[width=0.99\linewidth]{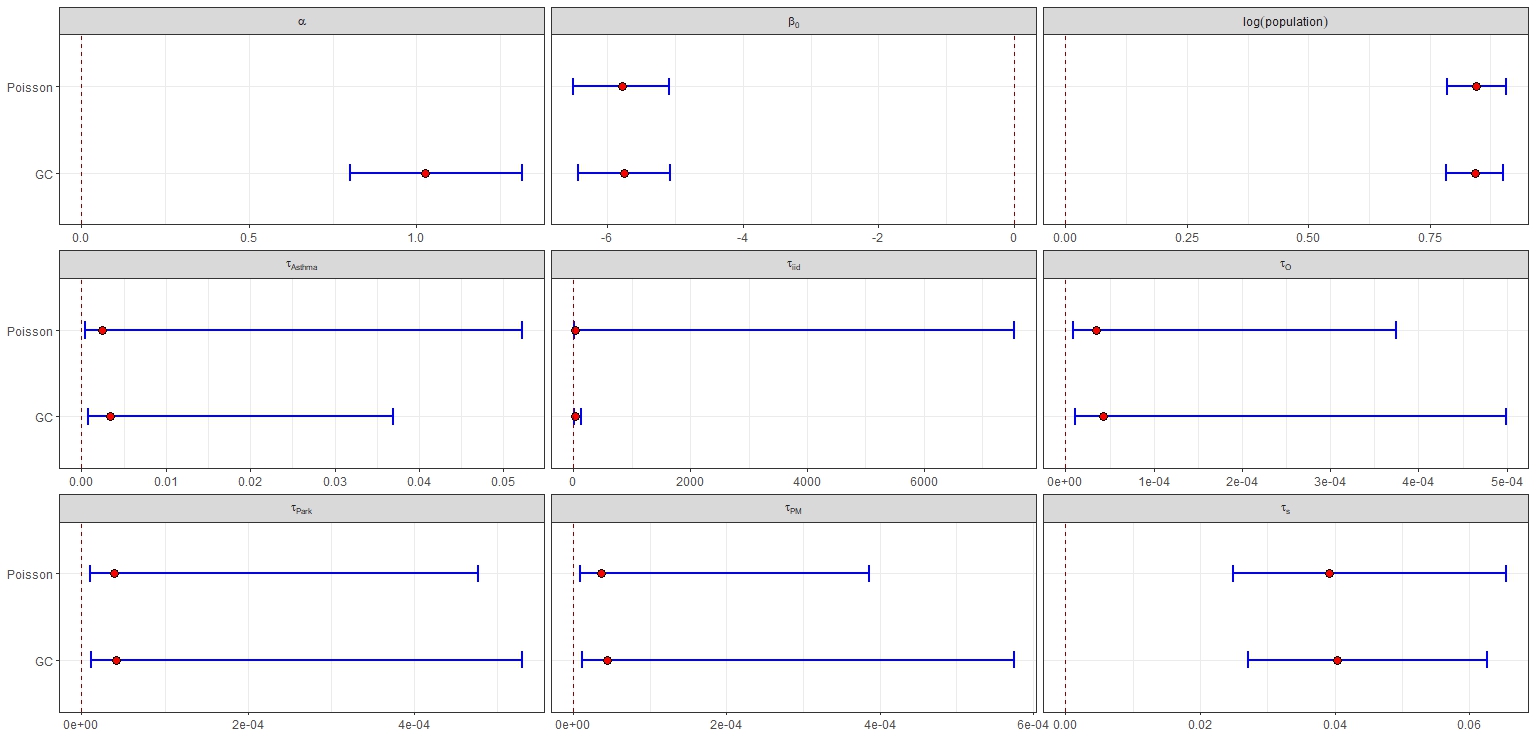}
    \caption{Lung and bronchus mortality data: Posterior mean estimates (red point) and 95\% credible intervals (blue line) of parameters for the various count models.}
    \label{fig:paramLung}
\end{figure}
Figure \ref{fig:paramLung} provides a summary of the outcomes from the top-performing models, namely the GC and Poisson models. The logarithmically transformed population variable possesses an estimated coefficient near one, indicating its efficacy in addressing variations in population size during relative risk modeling across different locations. This modification aims to normalize mortality rates for comparative analysis across various geographical locations. The logarithmic adjustment of population size serves as an offset term, facilitating the evaluation of relative risk estimations and allowing robust comparisons across various locales. Furthermore, the estimate of $\alpha$ exceeds one; nevertheless, its 95\% credible interval includes one. Consequently, the NB model seems insufficient for data analysis.
\begin{figure}
    \centering
    \begin{tabular}{cc}
          \includegraphics[width=0.48\linewidth]{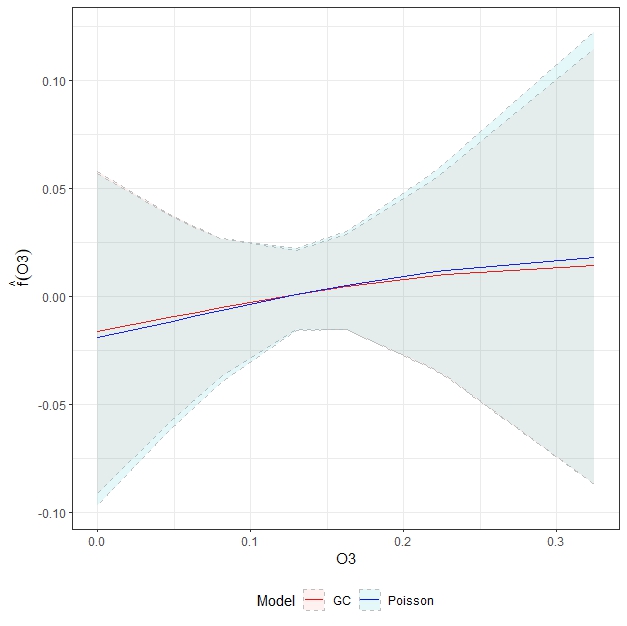}
   &     \includegraphics[width=0.48\linewidth]{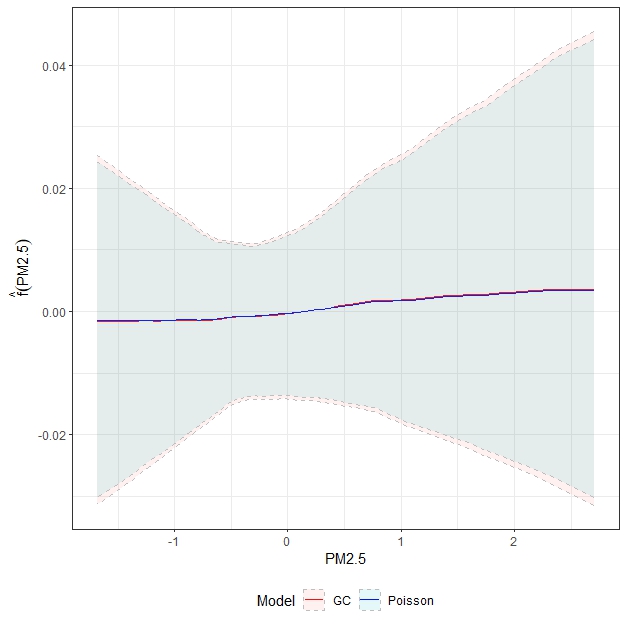}
 \\
         \includegraphics[width=0.48\linewidth]{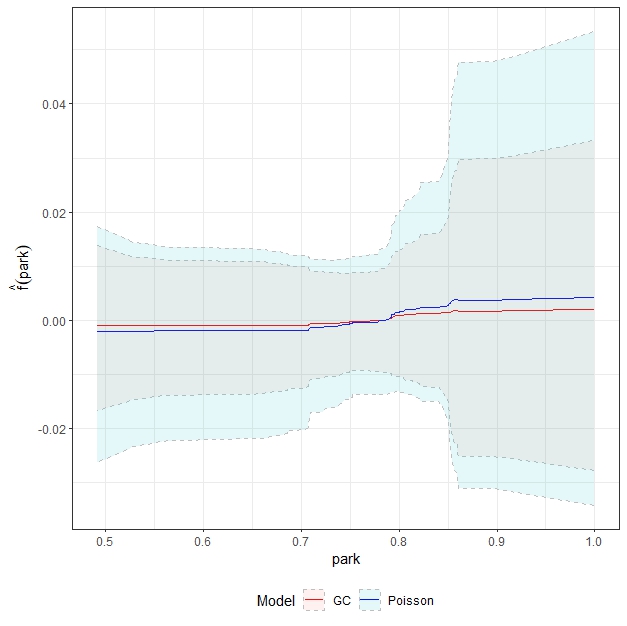}
    &     \includegraphics[width=0.48\linewidth]{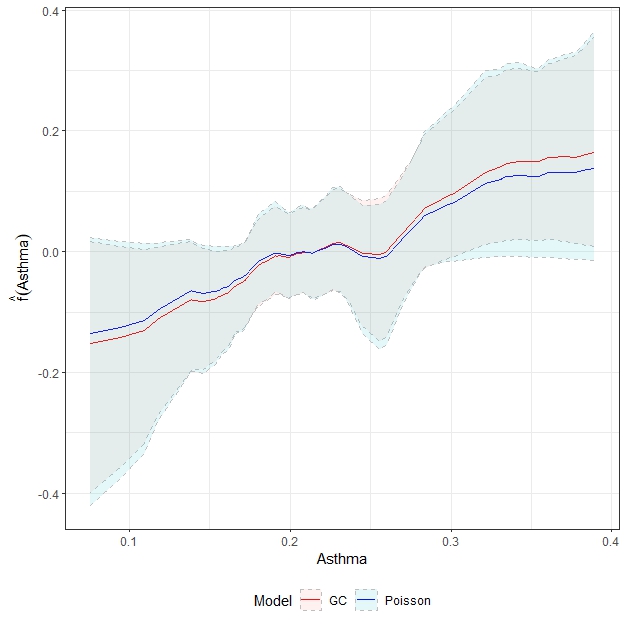}
    \end{tabular}
    \caption{Lung and bronchus mortality data: Posterior mean estimates (solid line) and 95\% credible intervals (dashed line) for the non-linear effects of ozone (first
row, left), PM2.5 (first row, right), Park (second row, left), and Asthma (second row, right) covariates under the GC and Poisson models.}
    \label{fig:ageL}
\end{figure}

Figure \ref{fig:ageL} illustrates the estimated smooth effects of ozone, PM2.5, park coverage, and asthma on lung and bronchus cancer mortality, accompanied with their respective 95\% credible intervals. The investigation reveals several significant observations.
The influence of ozone (\texttt{O3}) demonstrates a weak, marginally ascending trend correlated with the percentage of days over regulatory ozone thresholds, suggesting a moderate relationship with heightened cancer mortality risk. Likewise, PM2.5 has a similarly modest upward trend, indicating that extended exposure to fine particulate matter may modestly influence mortality rates, but with a constrained effect size.
The estimated effect of green space (\texttt{Park}) exhibits a little upward trend, approaching zero, signifying a limited correlation with lung and bronchus cancer mortality. Although more green space may affect cancer incidence more than mortality, the findings do not offer robust evidence to substantiate this theory. The restricted impact noted may potentially indicate the intricacies of green spaces' interactions with other environmental and demographic factors, or the inadequate extent of green areas in the examined regions to produce quantifiable health benefits.
The presence of asthma (\texttt{Asthma}) exhibits a significant positive correlation with death from lung and bronchus cancer. Individuals with asthma exhibit a significantly raised risk, apparently due to pre-existing respiratory vulnerabilities that intensify the detrimental impacts of environmental contaminants and other cancer progression risk factors.
The findings emphasize the differing levels of impact that environmental and health-related factors have on cancer mortality, underscoring the urgent necessity for focused public health measures and policies to reduce specific hazards.
\begin{figure}
    \centering
    \begin{tabular}{c}
          \includegraphics[width=0.80\linewidth]{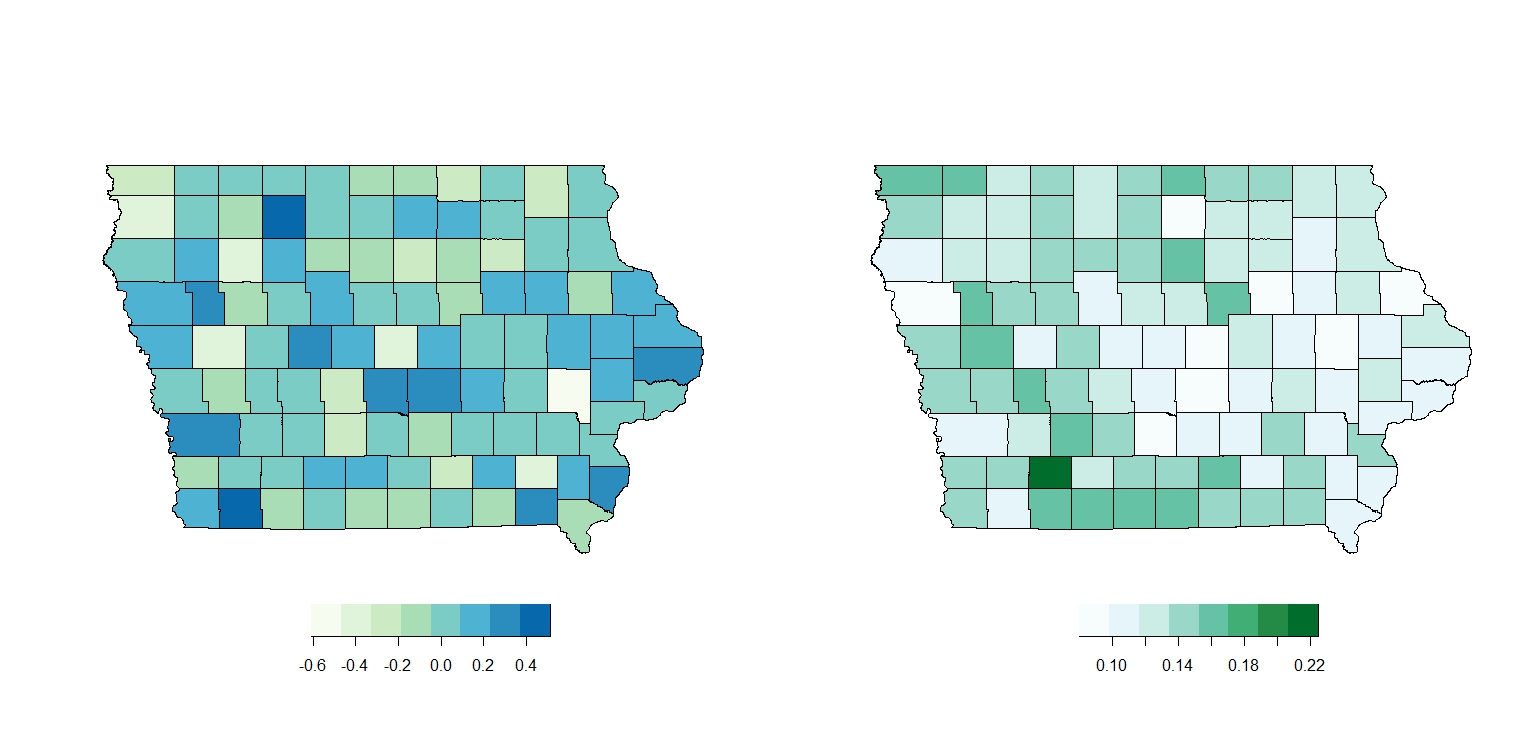}
 \\
         \includegraphics[width=0.8\linewidth]{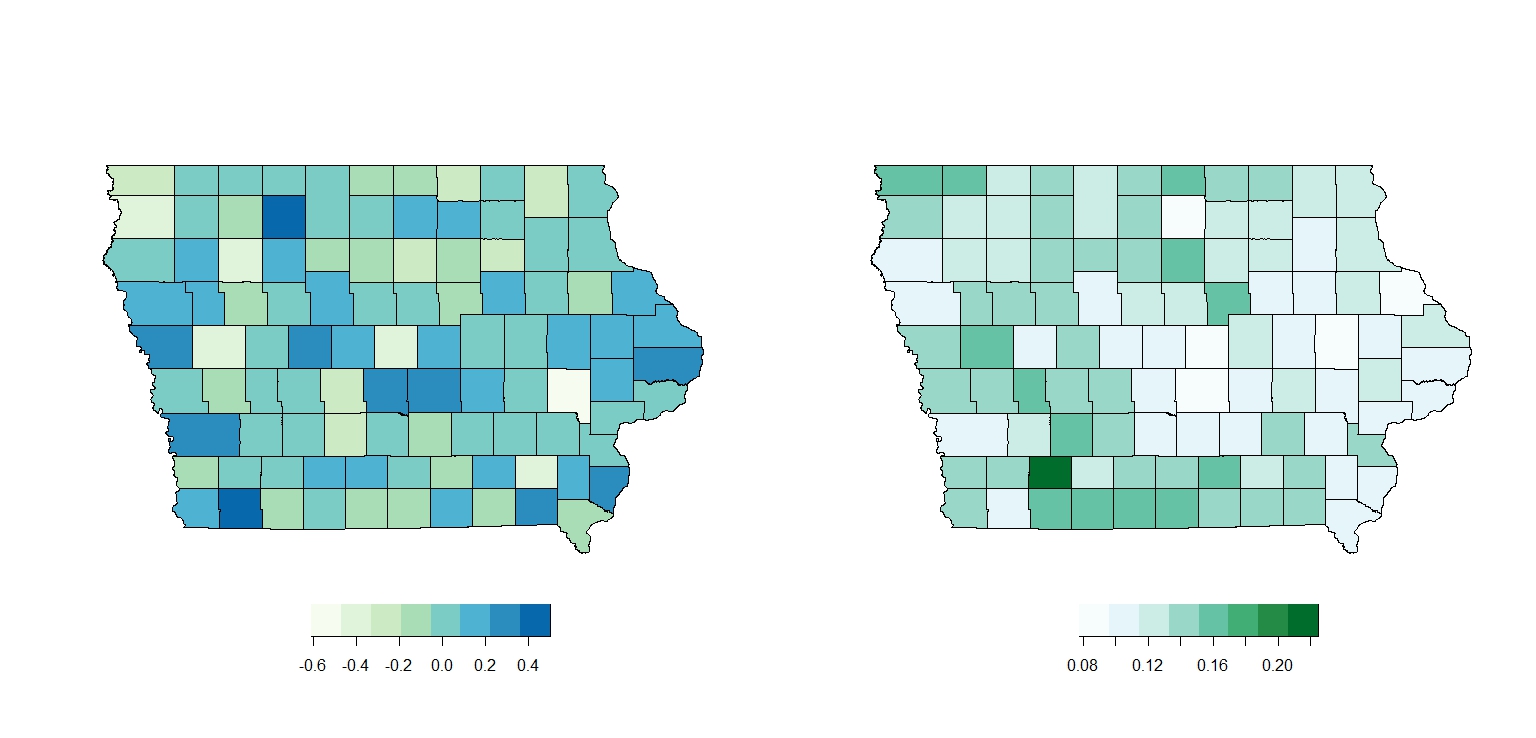}
    \end{tabular}
    \caption{Lung and bronchus mortality data: Posterior mean estimates (first column) of the total spatial effect and its corresponding standard deviation (second column), for the GC model (first row) and Poisson model (second row)}
    \label{fig:spatL}
\end{figure}

Figure \ref{fig:spatL} illustrates the overall posterior mean and standard deviation (SD) of the spatial effects for the GC and Poisson models. The posterior mean for the GC model reveals specific spatial patterns, indicating increased risks of lung and bronchus cancer mortality predominantly in the western and eastern areas of Iowa. The eastern regions have heightened spatial dependence, indicated by lower SD values, implying more uniform risk levels throughout these areas. In contrast, the western regions demonstrate increased variability in spatial impacts, indicating heightened uncertainty or heterogeneity in mortality risks. 
The variation is significantly greater in regions where predictions are extrapolated, which may be due to less information guiding the spatial impacts in these locations or to a lack of data availability. Notwithstanding these differences, the general geographical patterns and variability evident in the GC model largely correspond with those of the Poisson model, suggesting that both models effectively encapsulate analogous underlying spatial dependencies. The GC model offers a more refined comprehension of geographic heterogeneity, perhaps owing to its ability to integrate structured random effects and address dispersion in count data.
\subsection{Number of Days Prediction with Precipitation in Alberta}

Here, we consider a monthly climate summary dataset from multiple places throughout the Canadian province of Alberta in March 2024.
The following dataset contains values of various climatic parameters, including monthly averages and extremes of temperature, precipitation amounts, degree days, and sunshine hours. However, our focus is on the frequency of days with precipitation of at least $1.0$ mm, with some covariates containing the longitudinal and latitudinal,  minimum temperature, and total precipitation.
 These variables are used to analyze precipitation patterns and their impacts on the region's agricultural practices, water resource management, and efforts to adapt to climate change.

In summary, in the present study, the response variable is the number of days with precipitation of threshold $1.0$ mm, and the continuous covariates include total precipitation ($\mathtt{TP}$) and the lowest monthly minimum temperature ($\mathtt{Tn}$). These factors are critical for describing the intricate interactions that occur between temperature, precipitation, and other meteorological factors across Alberta's diverse terrain. Notably, there is no information regarding precipitation at several of the data points. The dataset is accessible via the link \url{https://climate.weather.gc.ca/prods_servs/cdn_climate_summary_e.html}.

\begin{figure}[h!]
    \centering
    \begin{tabular}{cc}
        \includegraphics[width=0.45\textwidth,height=7cm]{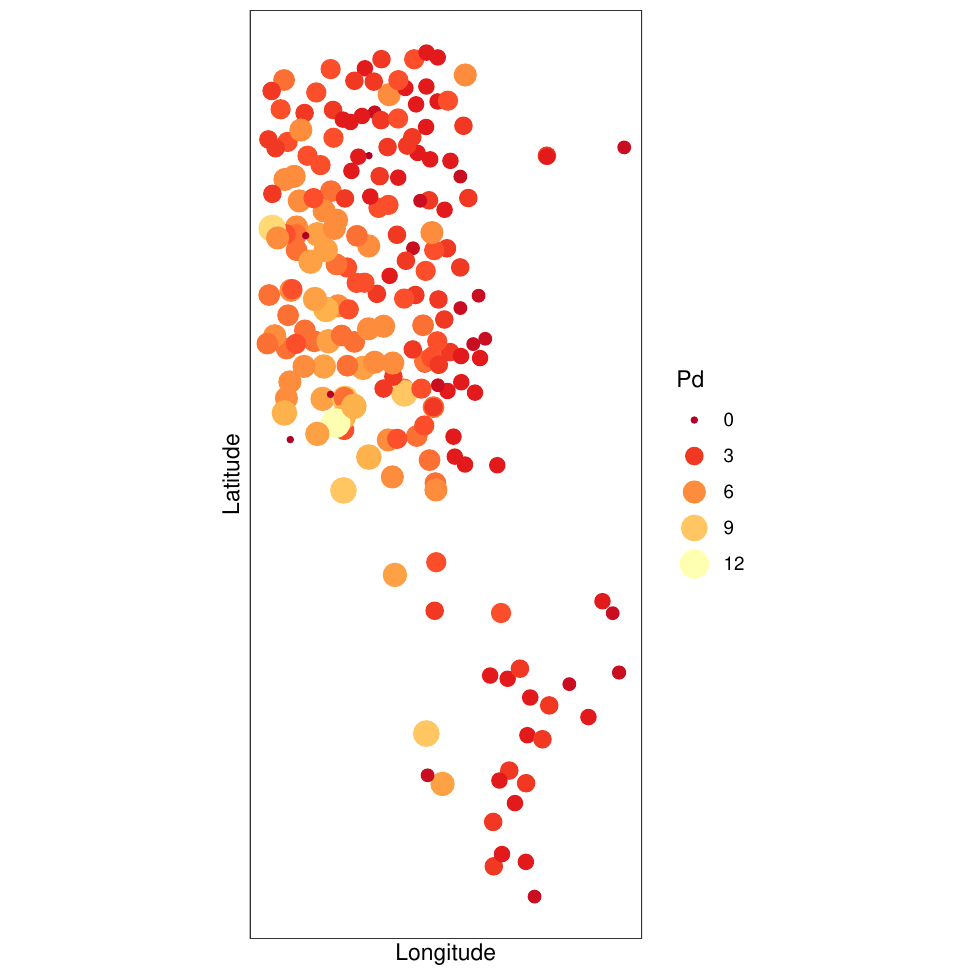} &
        \includegraphics[width=0.45\textwidth,height=7cm]{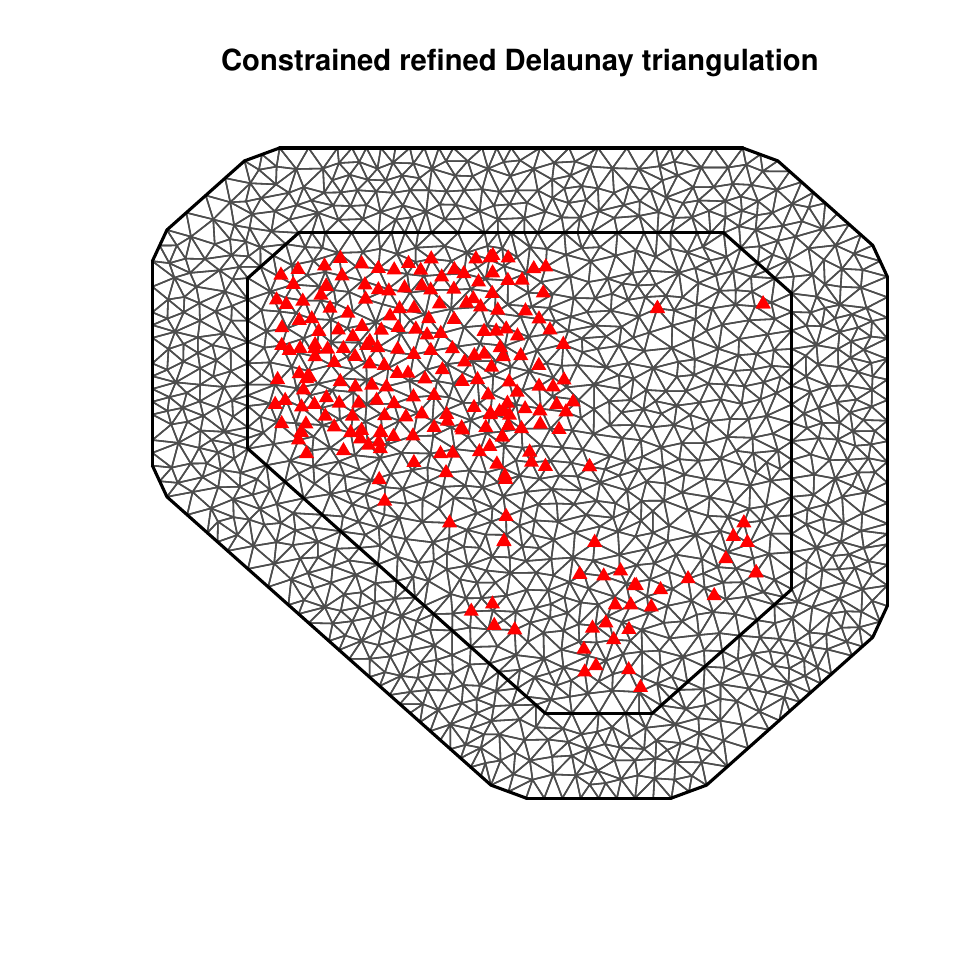}
    \end{tabular}
    \caption{Precipitation survey: sampling locations and the corresponding number of days with precipitation of $1.0$ mm or more (left), and triangular mesh built for TPS with the sampling locations (right).}
    \label{fig:datacanada}
\end{figure}

Our objective is to predict the frequency of days with precipitation of $1.0$ mm or above for unknown locations by applying the proposed Bayesian semiparametric spatial GC model to the precipitation dataset in Alberta. Considering spatial autocorrelation, this method aims to create and validate a prediction model that accurately reflects the complex interactions between the response variable and covariates. The model provides valuable insights into potential hotspots and regional patterns across Alberta by generating spatially explicit predictions for unobserved locations. 
Figure \ref{fig:datacanada} illustrates the spatial distribution of sampling sites alongside the recorded number of days with precipitation of at least $1.0$ mm. The response variable denotes a restricted count and the probability of encountering more than 13 days of precipitation is almost zero. Considering this empirical distribution, utilizing unbounded-count models, such as the proposed SPSGC model, is suitable, since the effective support of the data does not impose practical limitations that would render these modeling options invalid \citep{lord2005,winkelmann2008,hilbe2011,Cameron2013}.

In our analysis, the non-parametric predictor employed is as follows:
\begin{eqnarray*}
    \eta_i =  f_{Tn}(\mathtt{Tn}_i) + f_{TP}(\mathtt{TP}_i) + f_s(s_{1i}, s_{2i}),~~ i = 1, \ldots, 236,
\end{eqnarray*}
where $f_{Tn}(\mathtt{Tn}_i)$ and $f_{TP}(\mathtt{TP}_i)$ represent the unknown nonlinear effects of the lowest monthly minimum temperature and total precipitation, respectively. Additionally, $f_s(\cdot)$ denotes the spatial effect at a point with longitudinal coordinate $s_1$ and latitudinal coordinate $s_2$.

\subsubsection{Results of Precipitation Survey}
The model selection criteria, including DIC, WAIC, and LS are summarized in Table \ref{Tabdiccanada}. To evaluate the predictive performance of our model, we consider LS measurement.
The GC model demonstrates its strong predictive power and precise fit to the dataset by consistently outperforming the competition on all performance metrics. The ability of our proposed model to accurately predict consequences and precisely fit data demonstrates its successful performance.

\begin{table}[ht]
    \centering
    \caption{\label{Tabdiccanada}Precipitation survey: Model criteria based on computed DIC and WAIC for the proposed model and its alternatives.}
    \begin{tabular}{cccc}
        \hline
        Model & DIC & WAIC&LS  \\ 
        \hline
        GC & \textbf{651.63} & \textbf{656.10} & \textbf{364.72}  \\ 
        Poisson & 812.14 & 805.24&421.01  \\ 
        NB & 812.33 & 807.93&416.66  \\ 
        GP &809.37 &814.97&419.69 \\
        \hline
    \end{tabular}
\end{table}

\begin{figure}[h!]
    \centering
\includegraphics[width=0.9\textwidth,height=7cm]{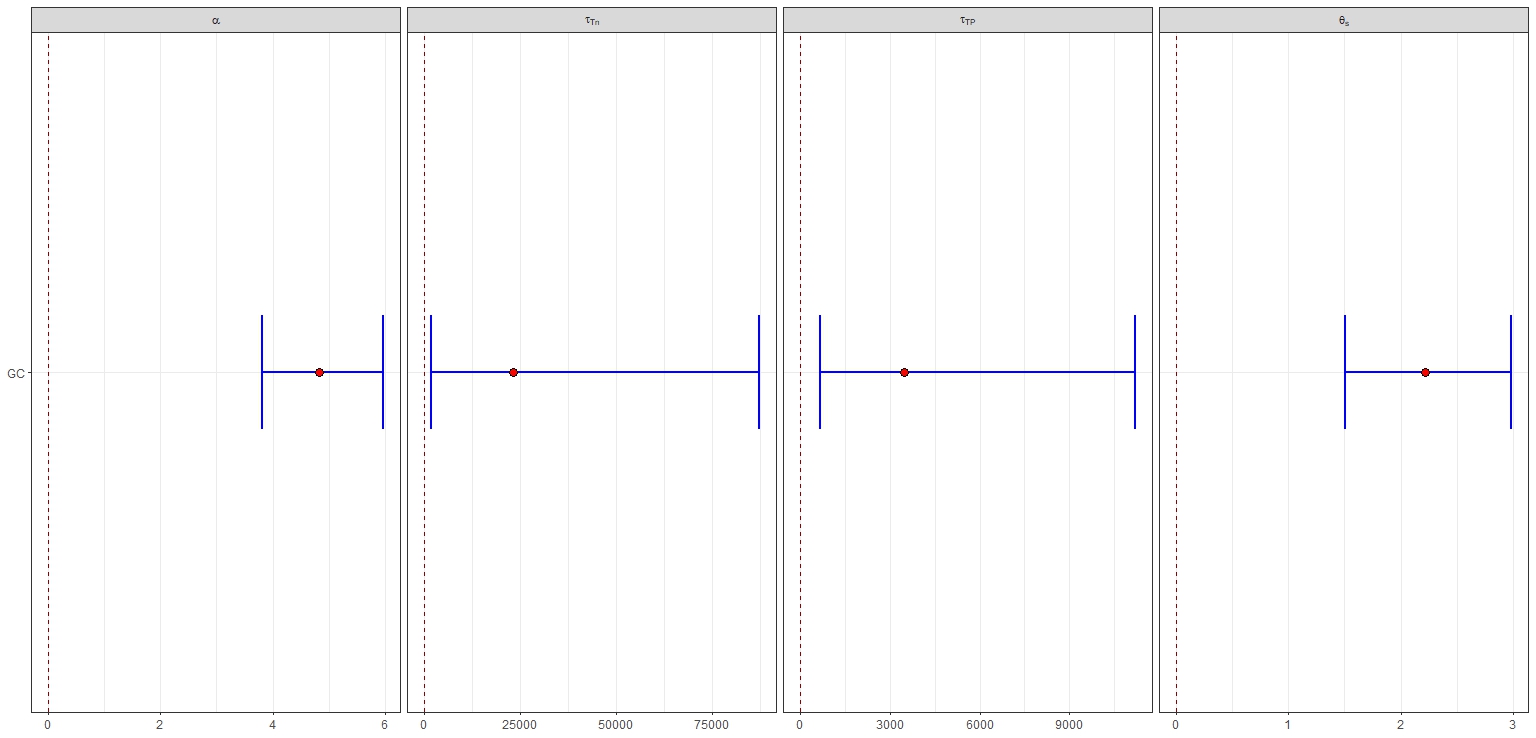}
    \caption{Precipitation survey: posterior mean estimates (point) and 95\% credible intervals (lines) of parameters and hyper-parameters for the proposed model and its alternatives.\label{figparamhatcanada}}
    \label{fig:est}
\end{figure}

The posterior mean estimates (points) and 95\% credible intervals (lines) for the precision of fixed and spatial effects, as well as the dispersion parameters for the different count models, are shown in Figure \ref{figparamhatcanada}. The leftmost column displays the estimated dispersion parameter for the GC model and the over-dispersion parameter for the NB model. The estimated value of $alpha$ in the GC model, which is much higher than four, demonstrates under-dispersion in the data. This is consistent with the results in Table \ref{Tabdiccanada}, revealing the GC model's better performance.

The nonlinear fixed effects precision parameter estimations show that the GC model has a larger impact, and the Poisson model has a higher degree of uncertainty than the other models. In addition, the spatial effect parameter shows decreased uncertainty and a stronger dependence in the GC model. This indicates that, compared to the other models, the GC model is better at capturing spatial dependency.

\begin{figure}[h!]
    \centering   
    \begin{tabular}{cc} 
\includegraphics[width=0.45\textwidth,height=7cm]{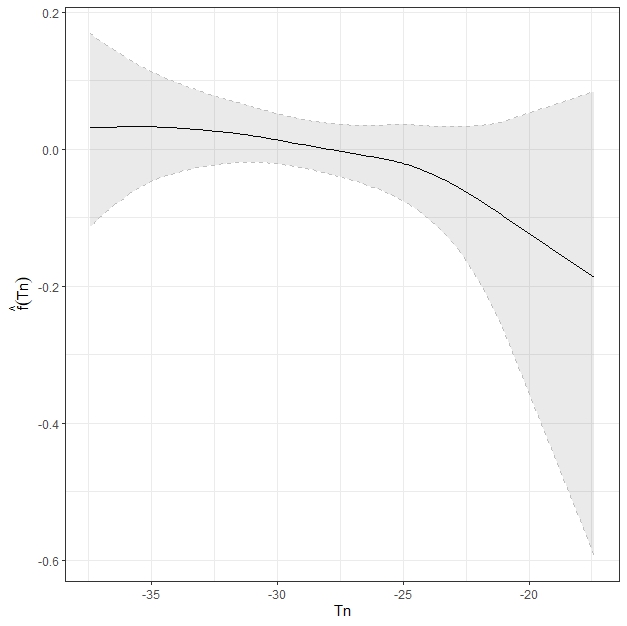} & \includegraphics[width=0.45\textwidth,height=7cm]{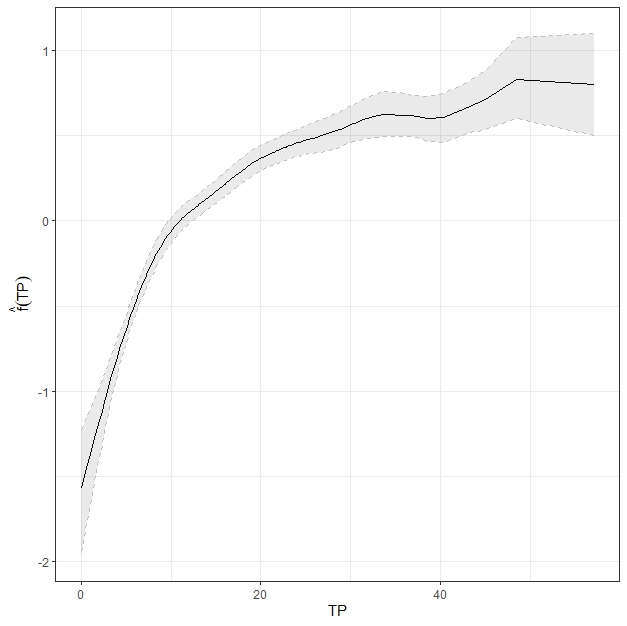}\\
    \end{tabular}
    \caption{Precipitation survey:  Fitted curve for nonlinear fixed effects including minimum temperature ( left), and total precipitation ( right) for the proposed model and its alternatives.}
    \label{fig:fcov}
\end{figure}

The fitted curves for the nonlinear effects of the lowest temperature and total precipitation are depicted in Figure \ref{fig:fcov}.

The lowest monthly minimum temperature has an adverse association with the probability of precipitation, suggesting that a rise in the minimum temperature reduces the likelihood of precipitation. On the other hand, there is a positive correlation between the overall amount of precipitation and the frequency of wet days, which makes sense.

\begin{figure}[h!]
    \centering      \includegraphics[width=0.9\textwidth,height=11cm]{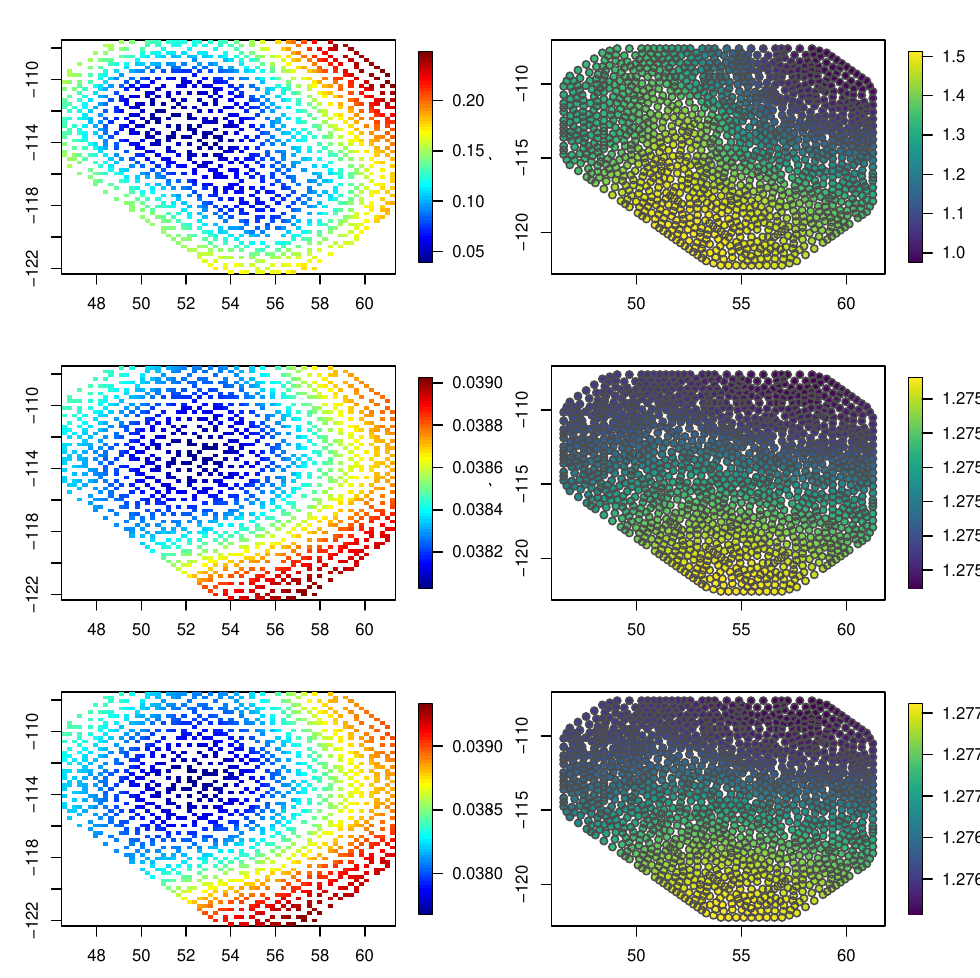}
\caption{ Precipitation survey:  first row: posterior mean of the spatial effect (left) and its standard deviation (right) of the GC model; second row: posterior mean (left) and standard deviation (right) of the Poisson model; and third row: posterior mean (left) and standard deviation (right) of the NB model.}

    \label{fig:fsAlberta}
\end{figure}

The posterior mean and standard deviation of the spatial effect are displayed in Figure \ref{fig:fsAlberta}. Whereas the GC model better depicts spatial dependencies, the uncertainty appears to be pretty stable across all models. The GC model's mapping findings show that areas far from the center are more likely to experience precipitation days, while areas closer to the center tend to experience less precipitation.

\begin{figure}[h!]
    \centering
    \begin{tabular}{cc}        \includegraphics[width=0.45\textwidth,height=7cm]{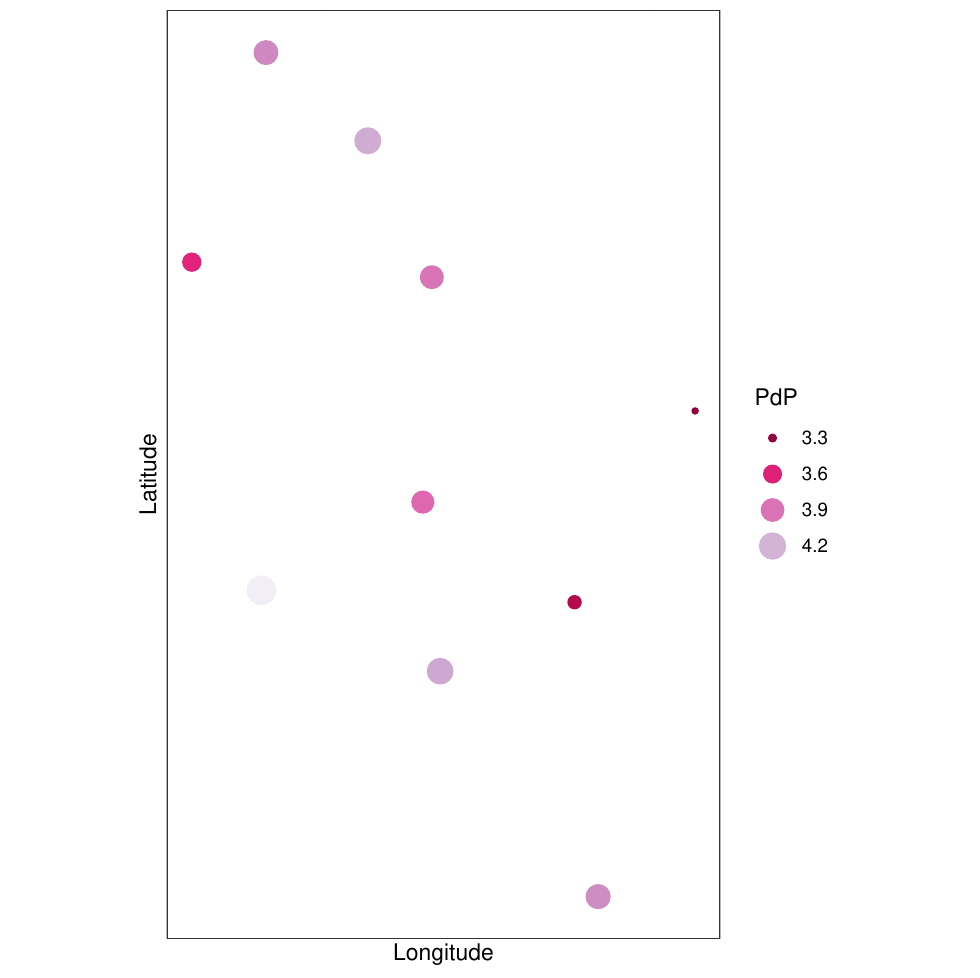} &
       \includegraphics[width=0.45\textwidth,height=7cm]{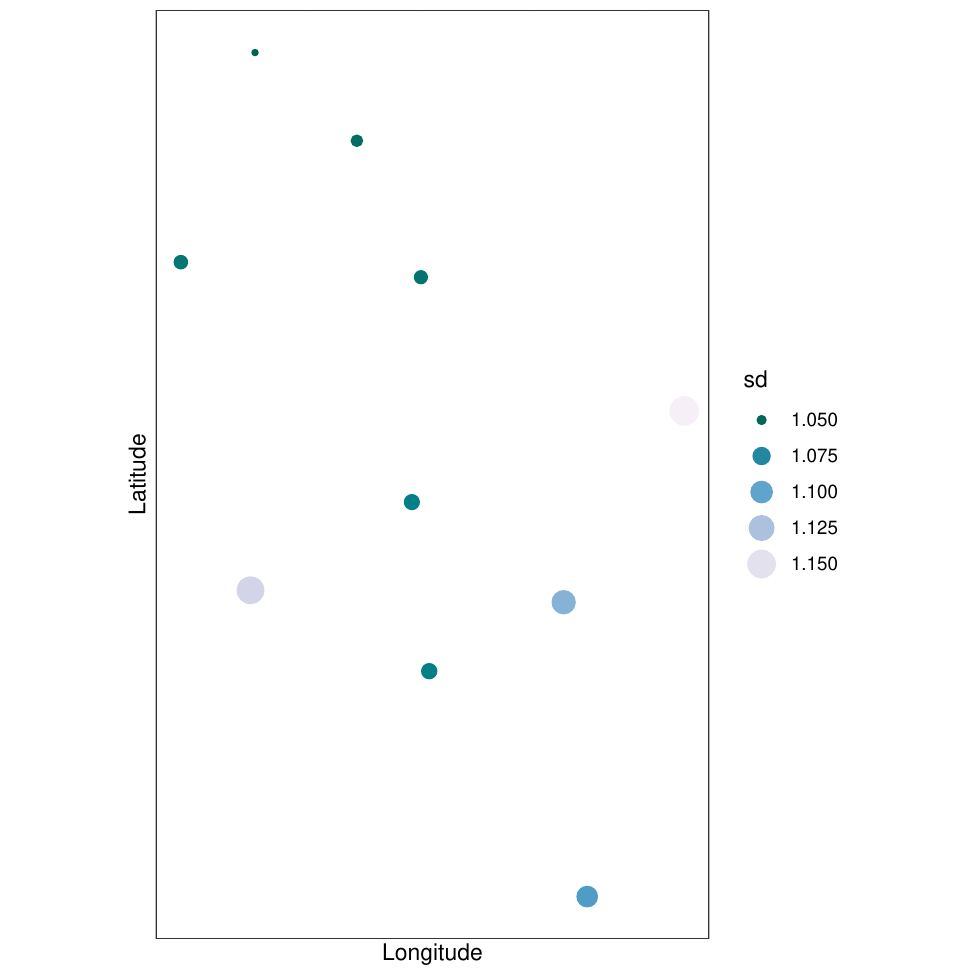} \\  
      \includegraphics[width=0.45\textwidth,height=7cm]{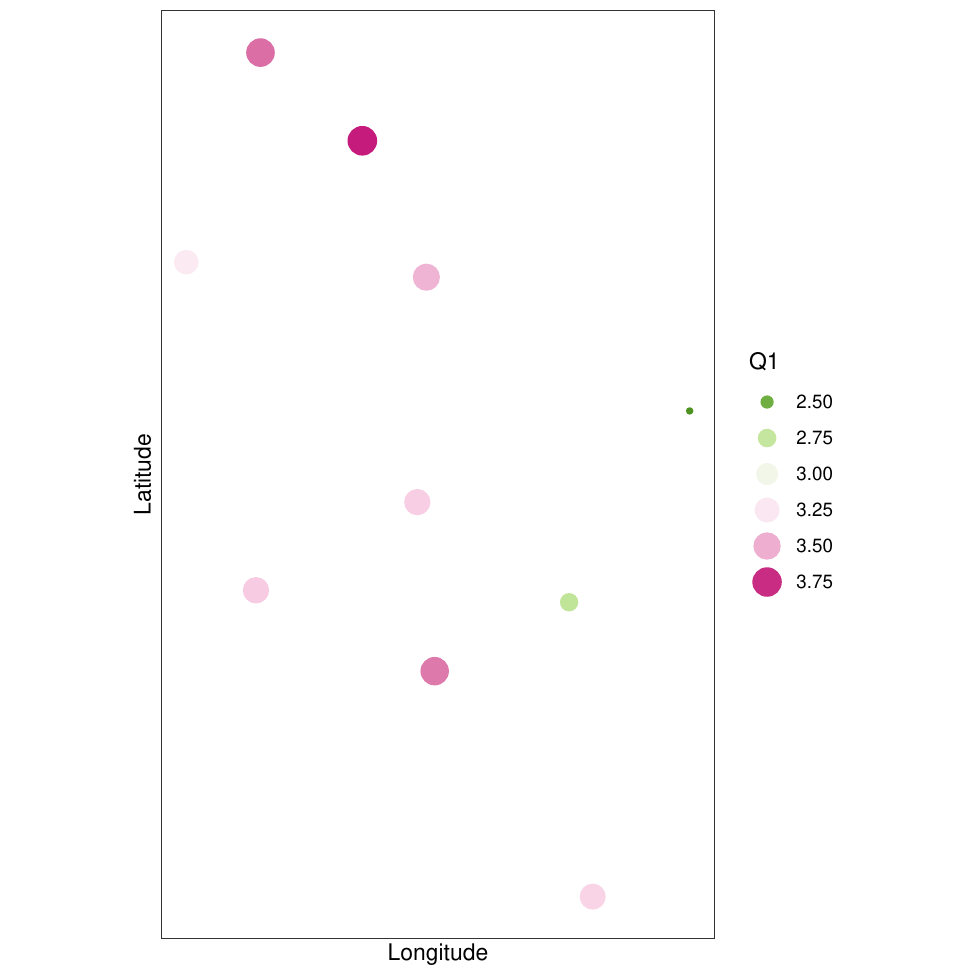} &
      \includegraphics[width=0.45\textwidth,height=7cm]{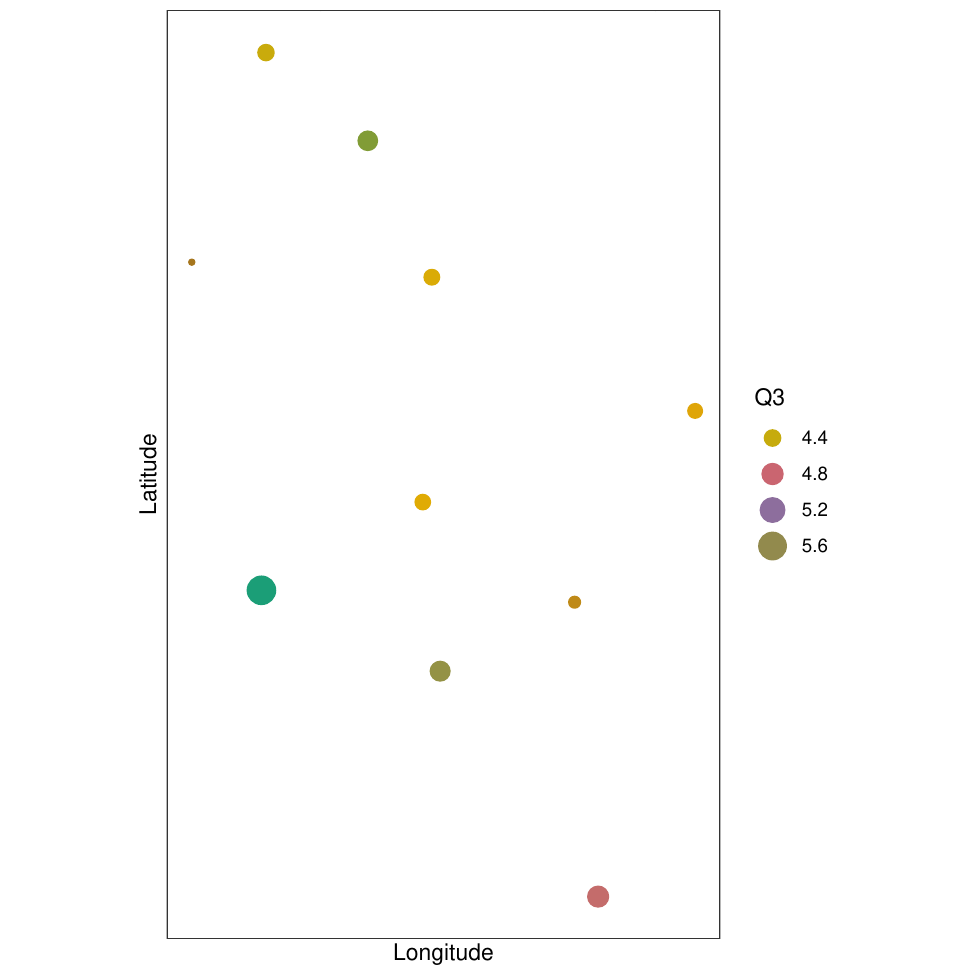}
    \end{tabular}
    \caption{Precipitation survey: The first row shows the posterior mean prediction (left) and its standard deviation (right); the second row presents the first (left) and third (right) quantiles of the prediction.}
    \label{fig:predictionAB}
\end{figure}

The recommended semi-parametric spatial GC model comes out to be the most influential at prediction after demonstrating the better fit of our model and going over the results shown in Table \ref{Tabdiccanada}. The dataset contains ten sites for which the number of days with a particular type of precipitation is missing. As a result, we predict how many days these unidentified locations will see at least $1.0$ mm of precipitation. The predicted posterior mean for unknown sites, along with the accompanying standard deviations and quantiles with probabilities of $0.025$ and $0.975$, are shown in Figure \ref{fig:predictionAB}. These statistics indicate that these regions generally have an average of three to four days of rainfall.

\section{Simulation Study}\label{Sec5}
We performed a simulation study to evaluate the proposed Bayesian model for evaluating various types of dispersed count data. The simulation settings are described, and the key results are displayed below.

\begin{figure}[h!]
    \centering
      \includegraphics[scale = 0.5]{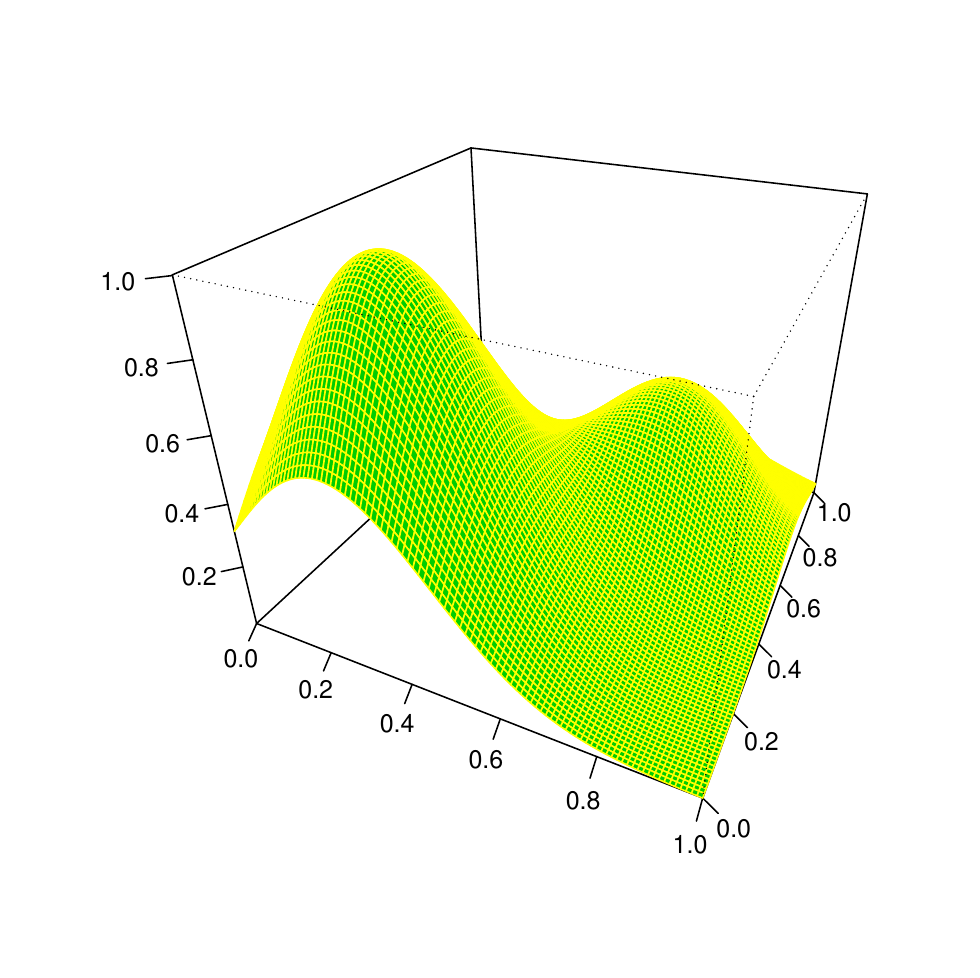} 
    \includegraphics[scale = 0.5]{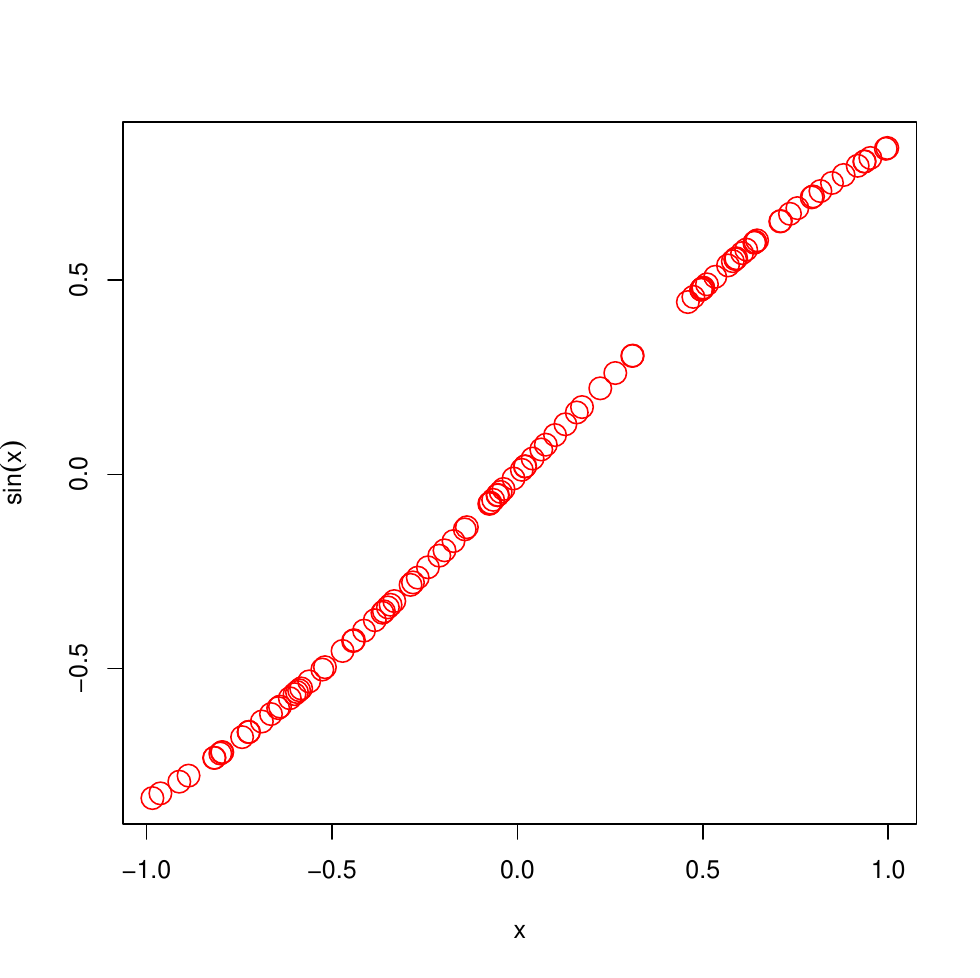}
    \caption{Simulation study: centered simulated functions $f_z(z)$ and $f_s(\bs)$. \label{fig:simfun}}
\end{figure}

We kept simulation settings fixed to allow for a consistent comparison in all scenarios of dispersion. We designed scenarios based on the main feature of the GC distribution, which represents:

\begin{itemize} 
\item[\textbf{1}:] Under-dispersion situation with consideration $\alpha = 1.5$
\item[\textbf{2}:] Equivalent-dispersion situation with setting $\alpha = 1$
\item[\textbf{3}:] Over-dispersion situation with choosing $\alpha = 0.4$. 
\end{itemize} 
As building blocks for structure additive model specifications, we considered one-variable and bivariable  non-linear forms for functional effect $f_z(\cdot)$ $f_s(\cdot)$, visualized in Figure \ref{fig:simfun}, as 
\begin{eqnarray}
\label{fe1} f_z(z)&=&\sin(z)\cr
f_s(s_1,s_2) &=& \exp((-(s_1 - 0.25)^2 - (s_2 - 0.25)^2)/0.1) + \cr
        &~~&~~~~~(0.5*\exp((-(s_1 - 0.7)^2 - (s_2 - 0.7)^2)/0.07))
\end{eqnarray}
with generating covariate from the uniform distribution ${\rm U}(-1,1)$. We also chose the sample size of 400 with a $20\times 20$ regular grid.
Finally, we simulated responses from the proposed model as
\begin{eqnarray*}
Y\mid z \sim {\rm GC}\left(\alpha, \alpha\exp\left( f_z\left(z\right)+f_s(s_1,s_2)\right)\right).
\end{eqnarray*} 

We conducted a comprehensive analysis to assess the effectiveness of the suggested model across various dispersion scenarios and count models, using multiple criteria. To demonstrate the effectiveness of the estimators for univariate or bivariate functional effects, we computed the  Mean Squared Errors (MSE) defined as  
\begin{eqnarray*} 
{\rm MSE}(\hat{f}_i)&=&\left(\frac{1}{R}\sum_{r=1}^{R}(\hat{f}_{ir}-f_{ir})^2\right),~~~~r=1,\ldots,R,~~ i=1,\ldots,n,
\end{eqnarray*}
where $n$ denotes the number of observations and $R$ is the number of replications.  For the dispersion parameter, we report the  Squared Errors (SE) and  as
 \begin{eqnarray*} 
{\rm SE}(\hat{\alpha}_r)&=&\left(\hat{\alpha}_{r}-\alpha_{r}\right)^2,~~~~~~r=1,\ldots,R.
\end{eqnarray*}
For each scenario, ${\rm R} = 100$ is used to compute  ${\rm SE}$, and corresponding DIC and WAIC criteria as model selection and predictive power measurements.
\subsection*{Results of simulation study}
\begin{figure}[h!]
    \centering
      \includegraphics[scale = 0.7]{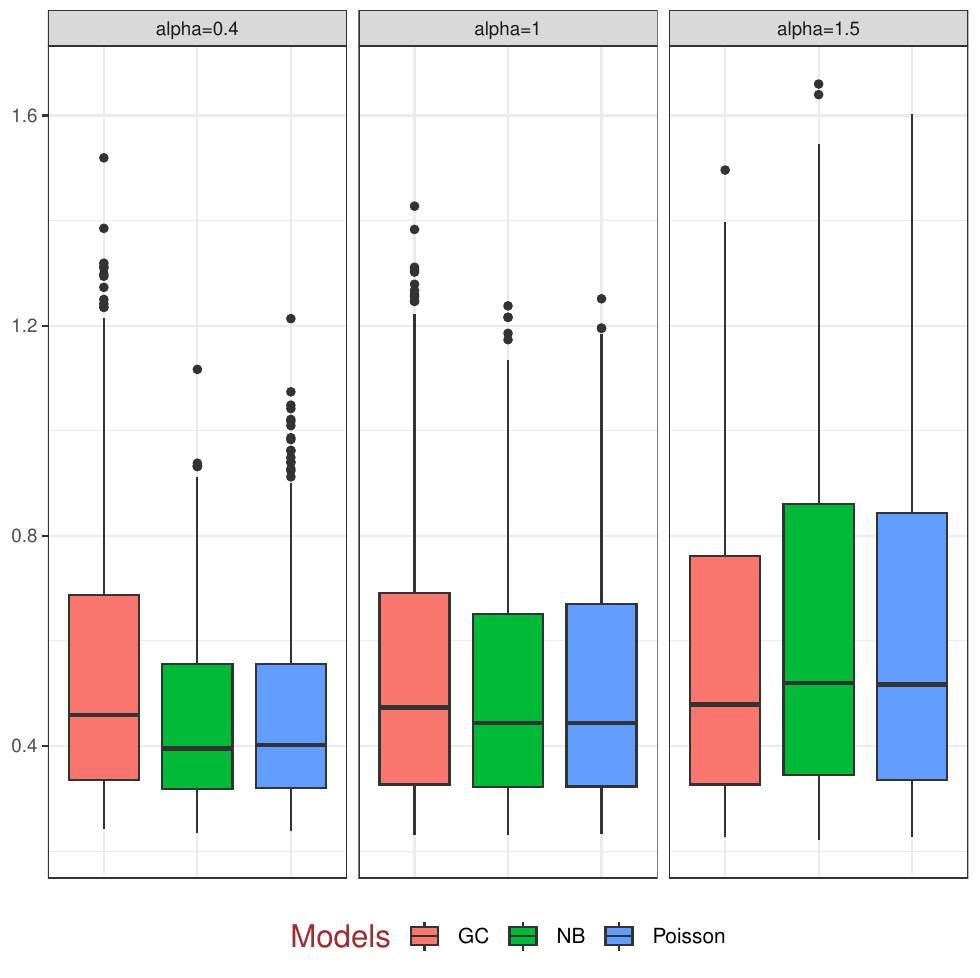}
    \caption{Simulation study: The boxplots of  ${\rm MSE}$ of $f_z(z)=\sin(z)$ for different models as well as dispersion scenarios.}
    \label{figmsefz}
\end{figure}

\begin{figure}[h!]
    \centering 
      \begin{tabular}{ccc}
      $\alpha=0.4$&$\alpha=1$&$\alpha=1.5$\\
      \includegraphics[width=0.3\textwidth,height=5cm]{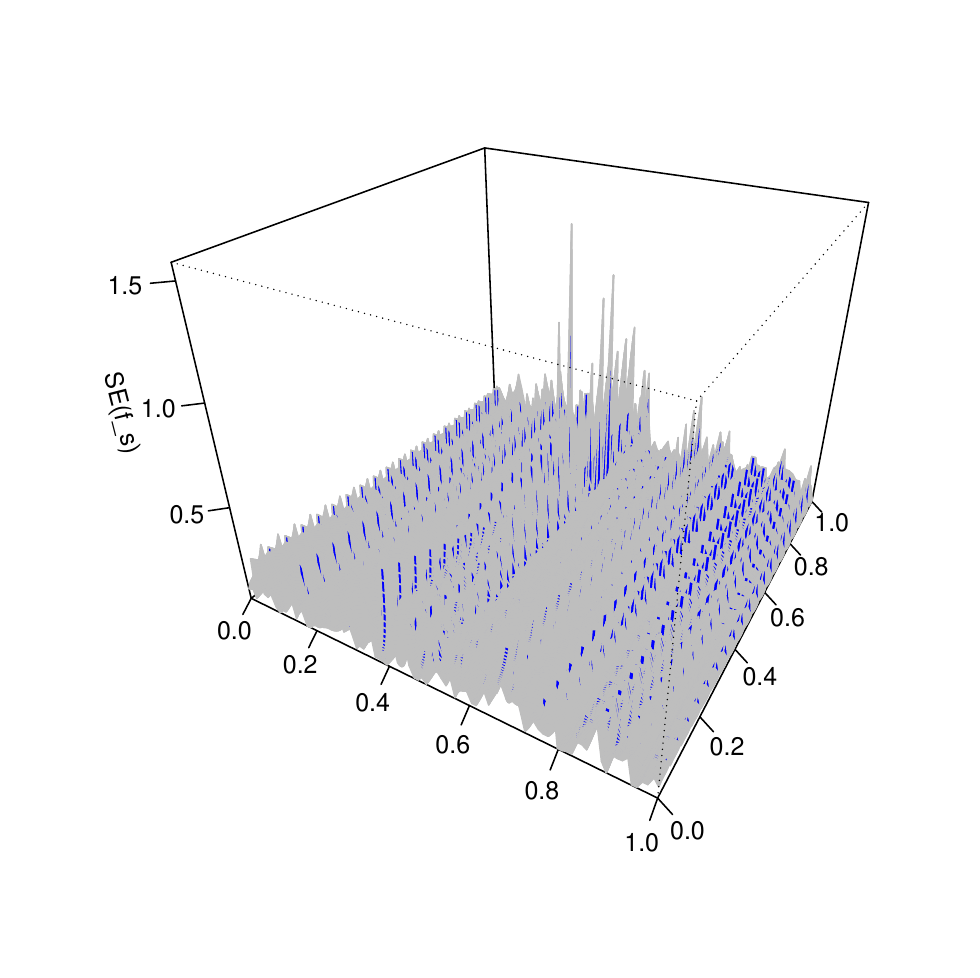} &
    \includegraphics[width=0.3\textwidth,height=5cm]{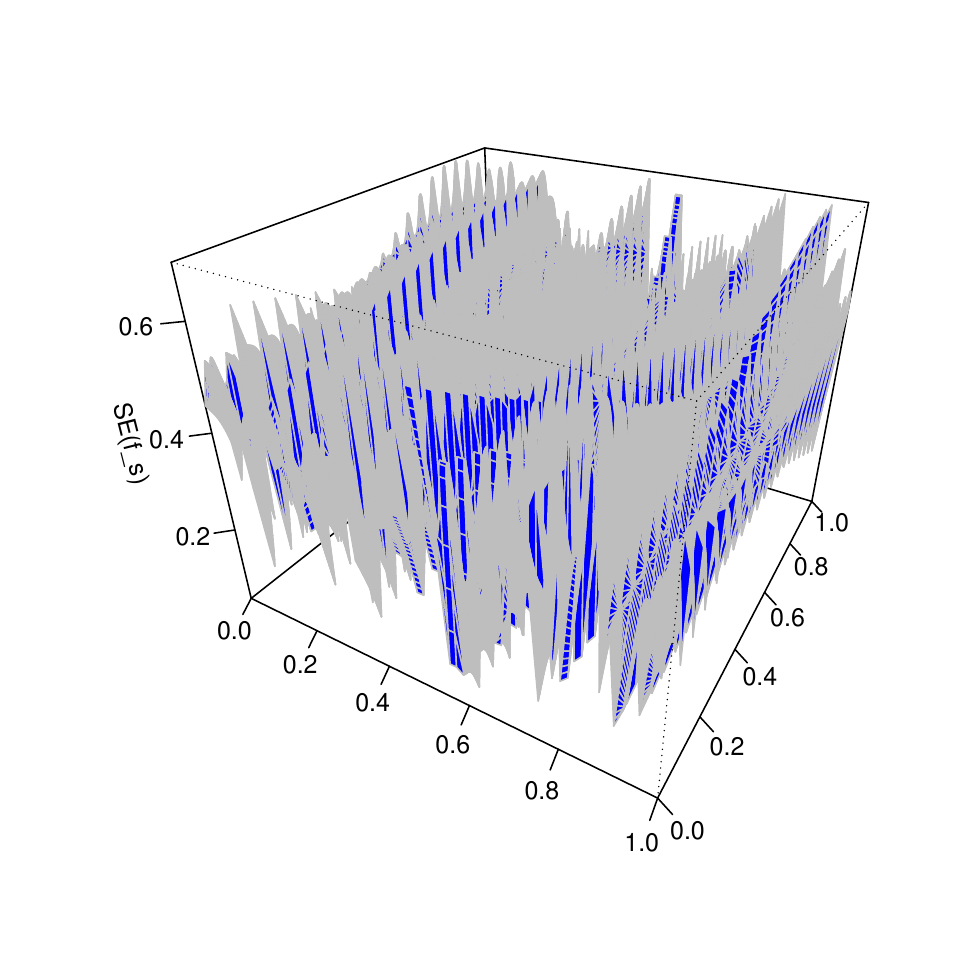}&
    \includegraphics[width=0.3\textwidth,height=5cm]{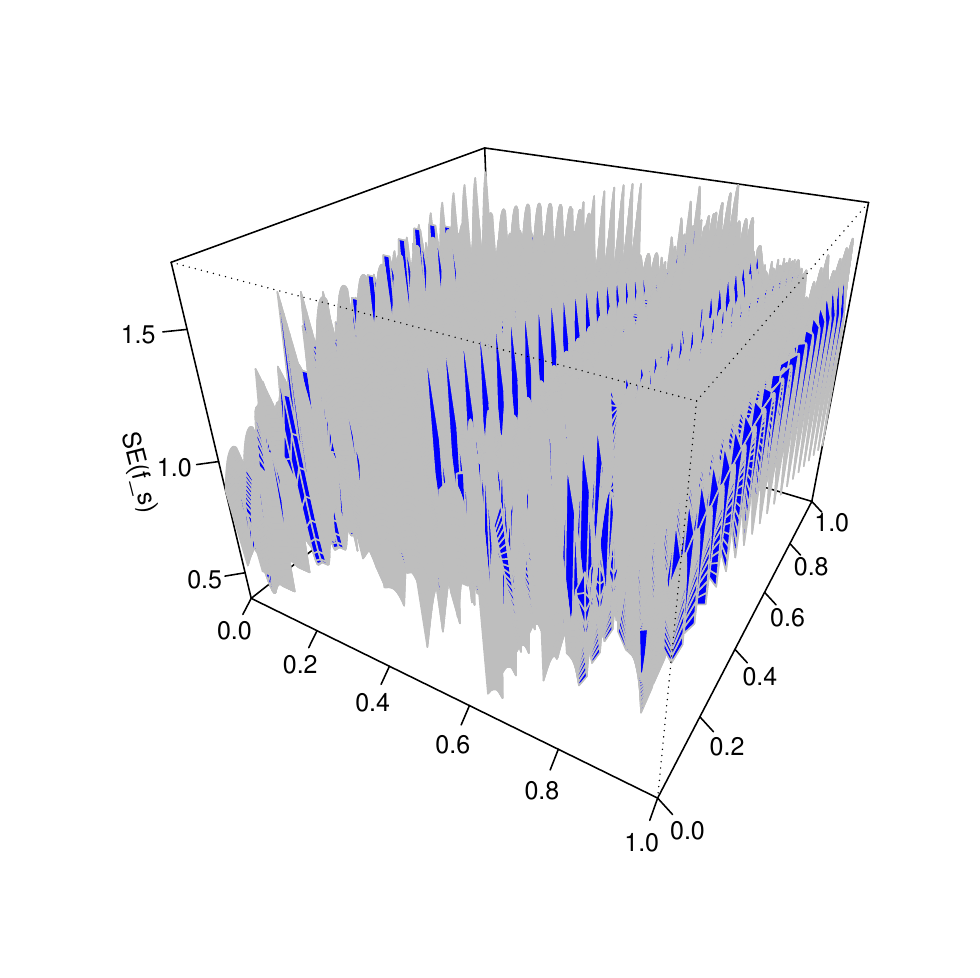}\\  
    \includegraphics[width=0.3\textwidth,height=5cm]{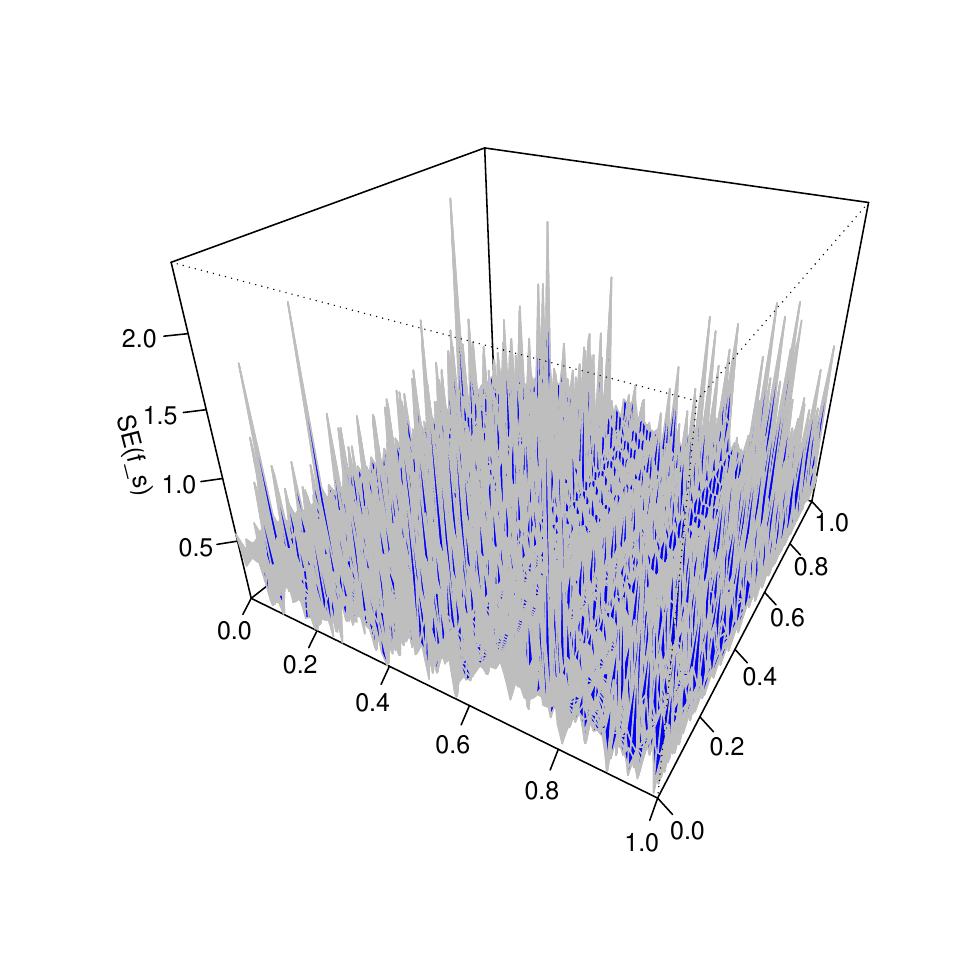} &
    \includegraphics[width=0.3\textwidth,height=5cm]{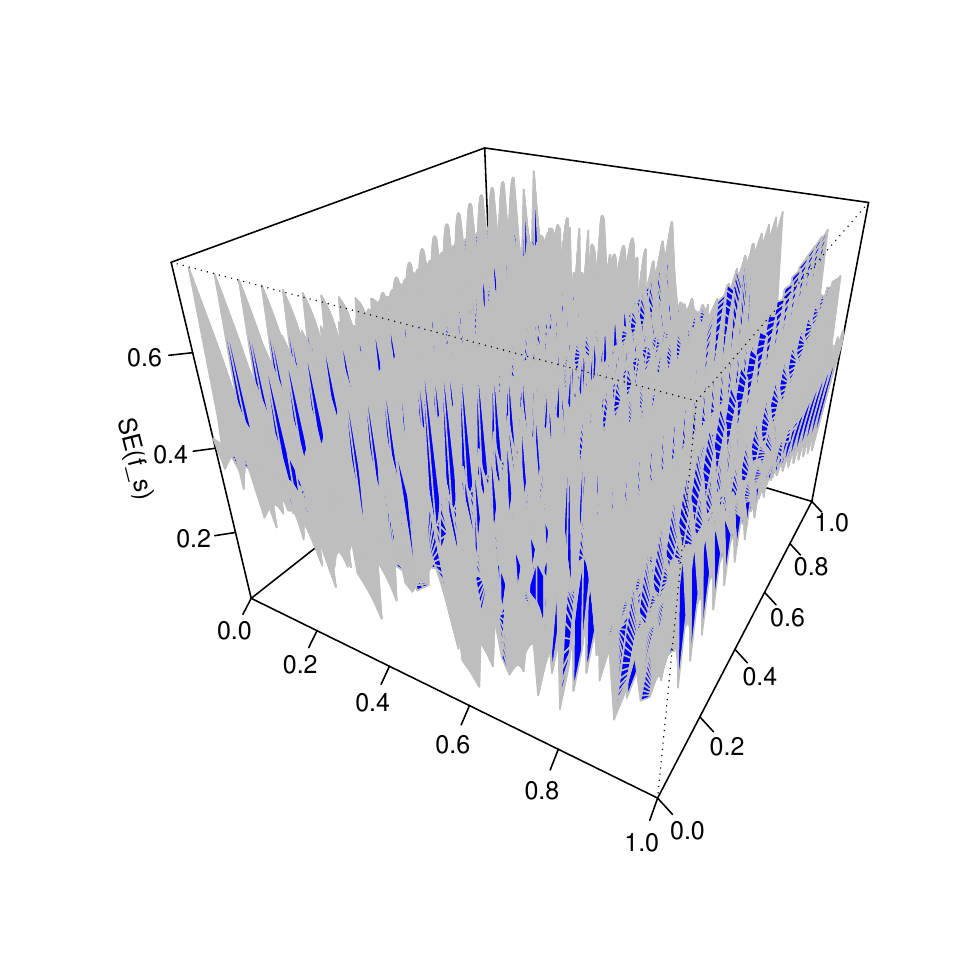}&
    \includegraphics[width=0.3\textwidth,height=5cm]{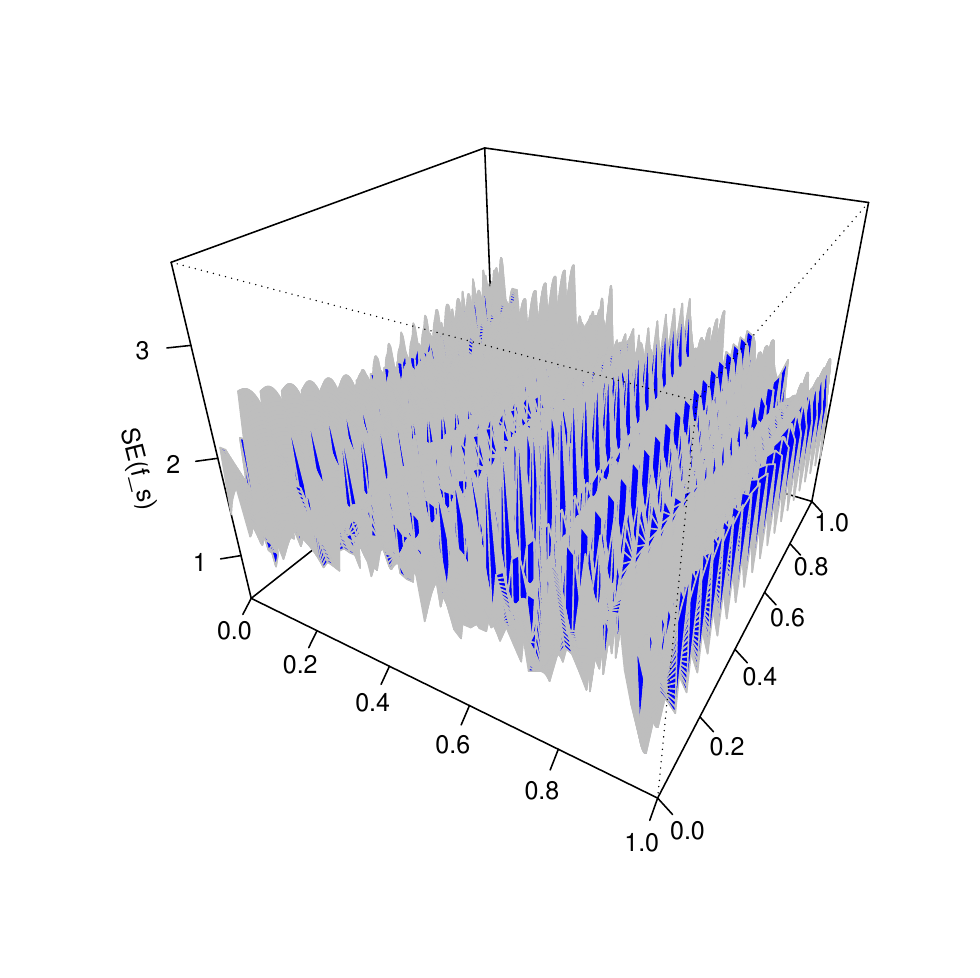}\\
        \includegraphics[width=0.3\textwidth,height=5cm]{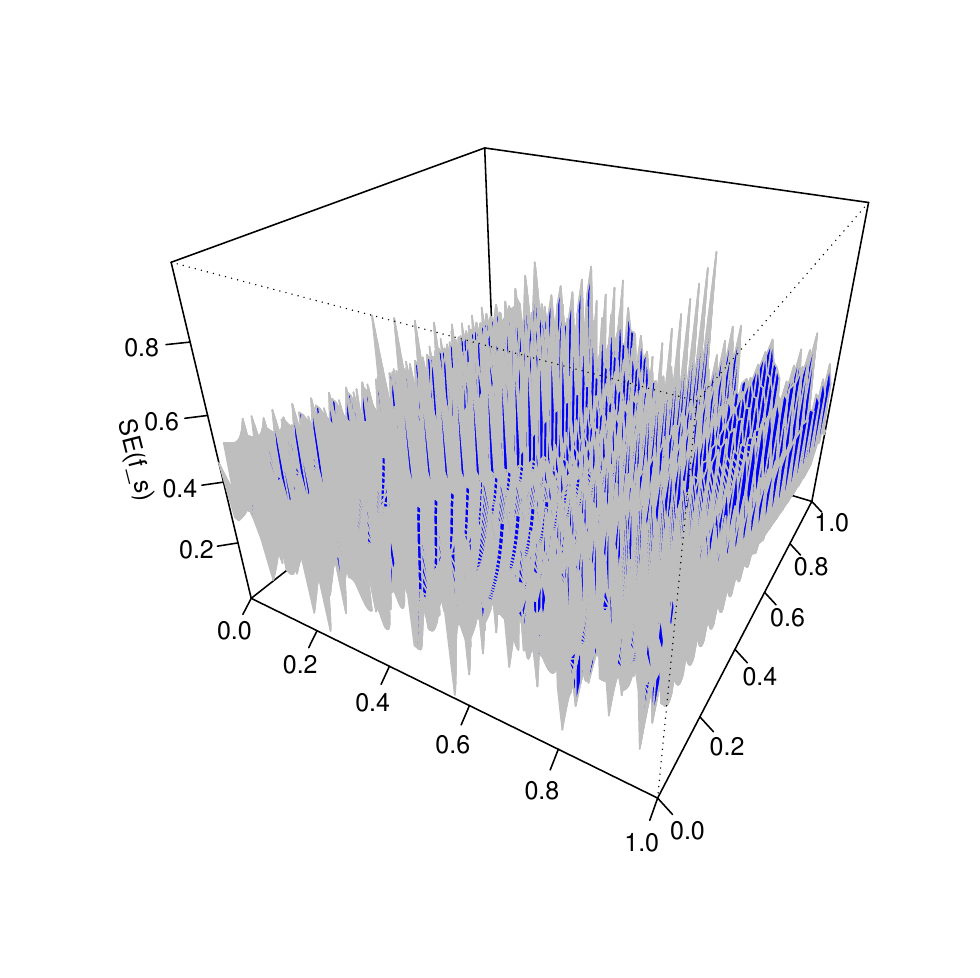} &
    \includegraphics[width=0.3\textwidth,height=5cm]{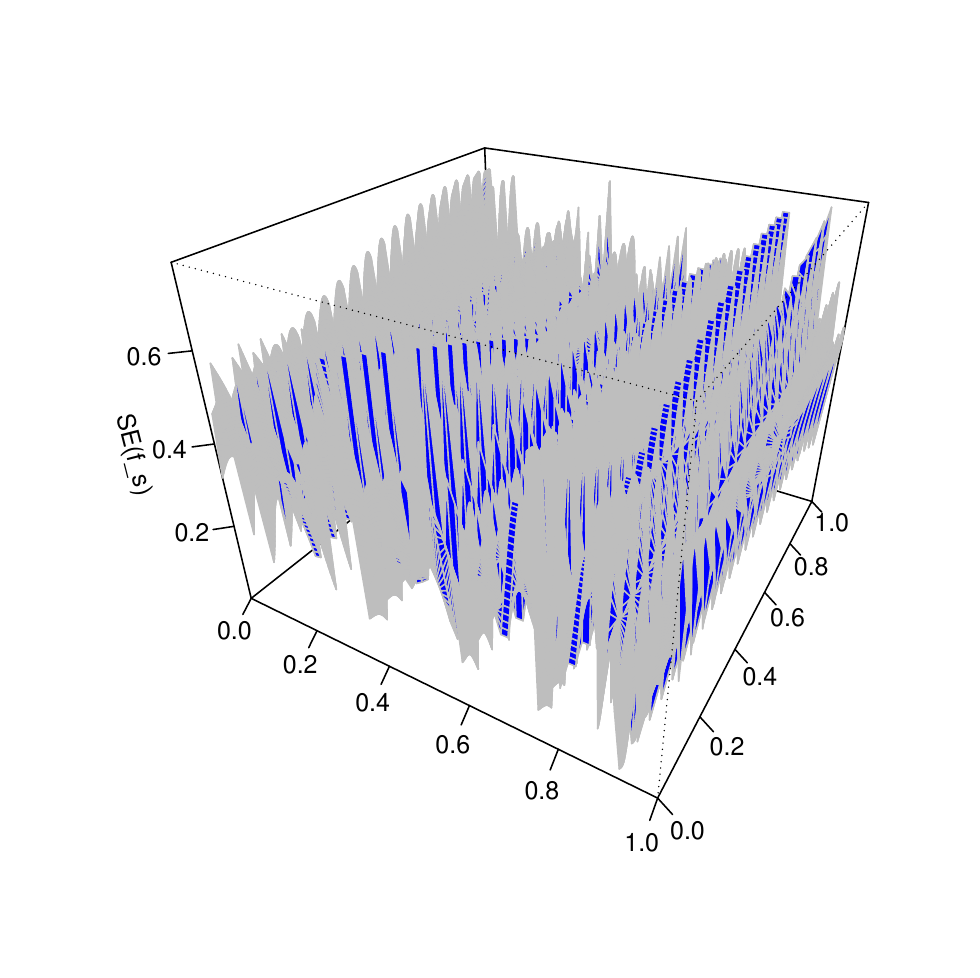}&
    \includegraphics[width=0.3\textwidth,height=5cm]{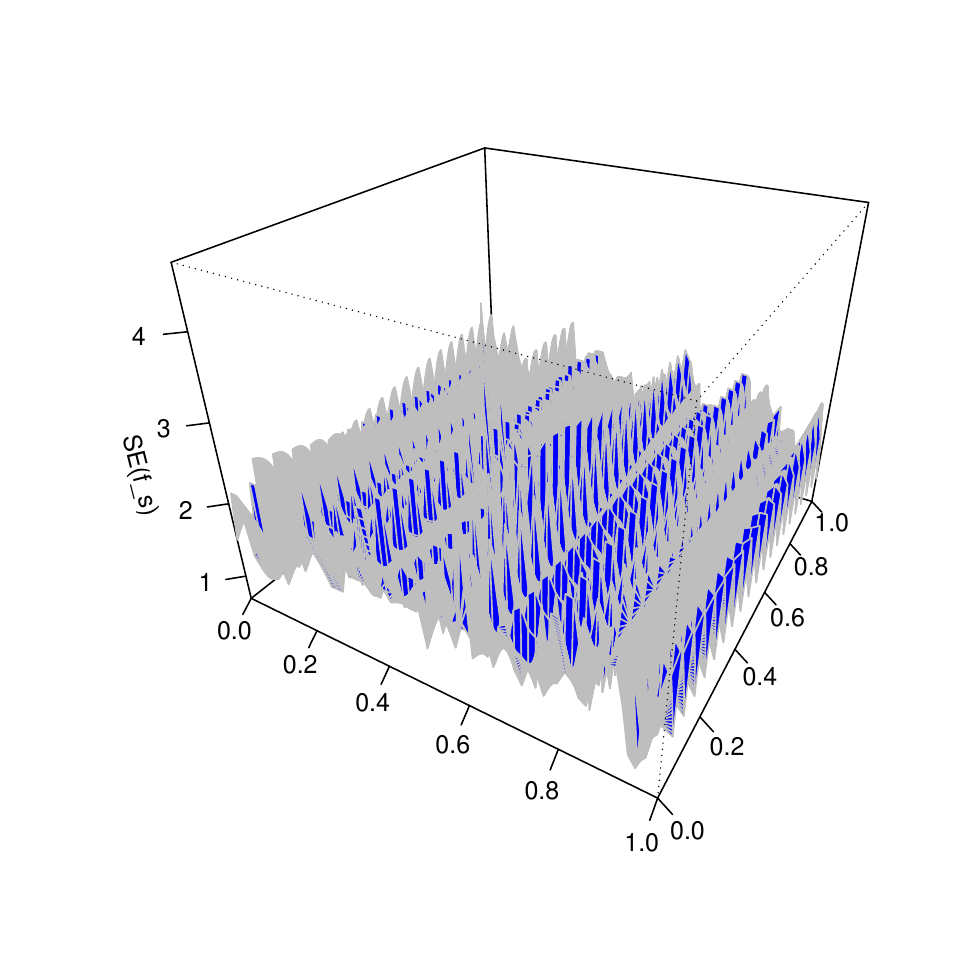}\\
\end{tabular}
    \caption{Simulation study: The perspective of  ${\rm MSE}$ of $f_s(s_1, s_2)$ for different models, including GC (first row), Poisson (second row), and NB (third row), as well as dispersion scenarios.}
    \label{figmsefs}
\end{figure}

\begin{figure}[h!]
    \centering
 \includegraphics[width=0.45\textwidth,height=7cm]{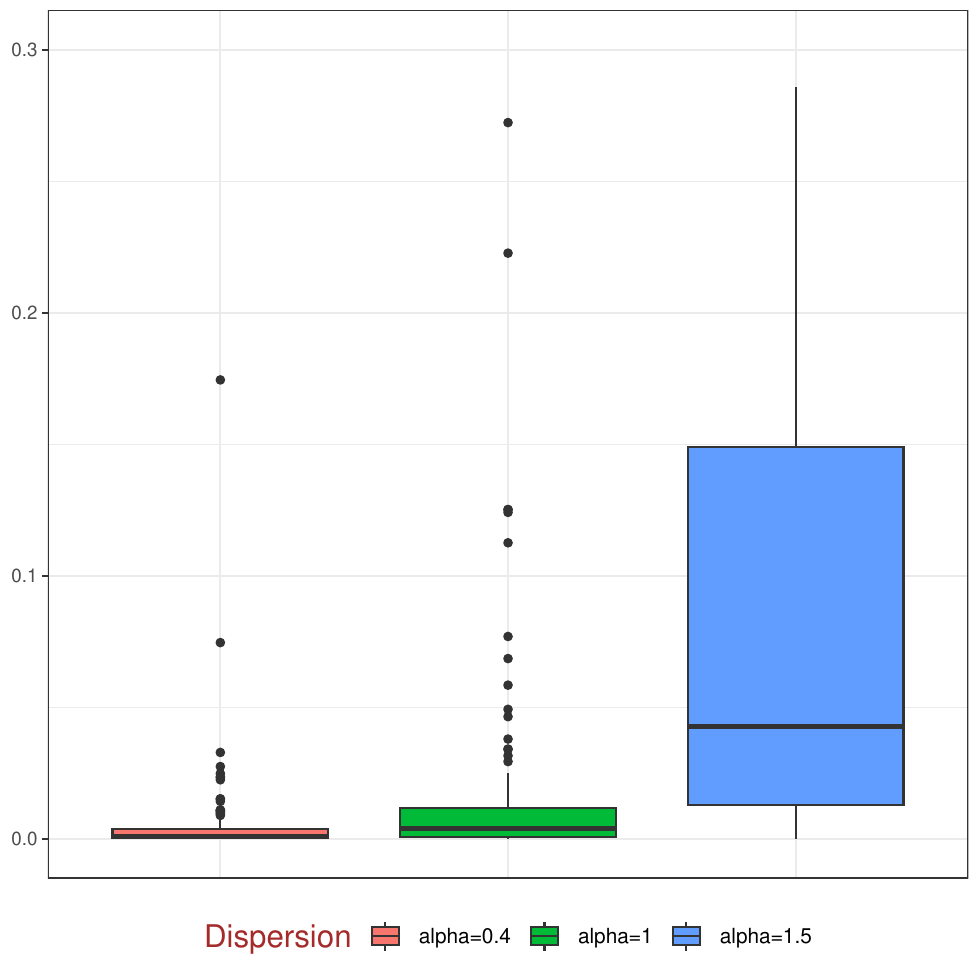}            \includegraphics[width=0.45\textwidth,height=7cm]{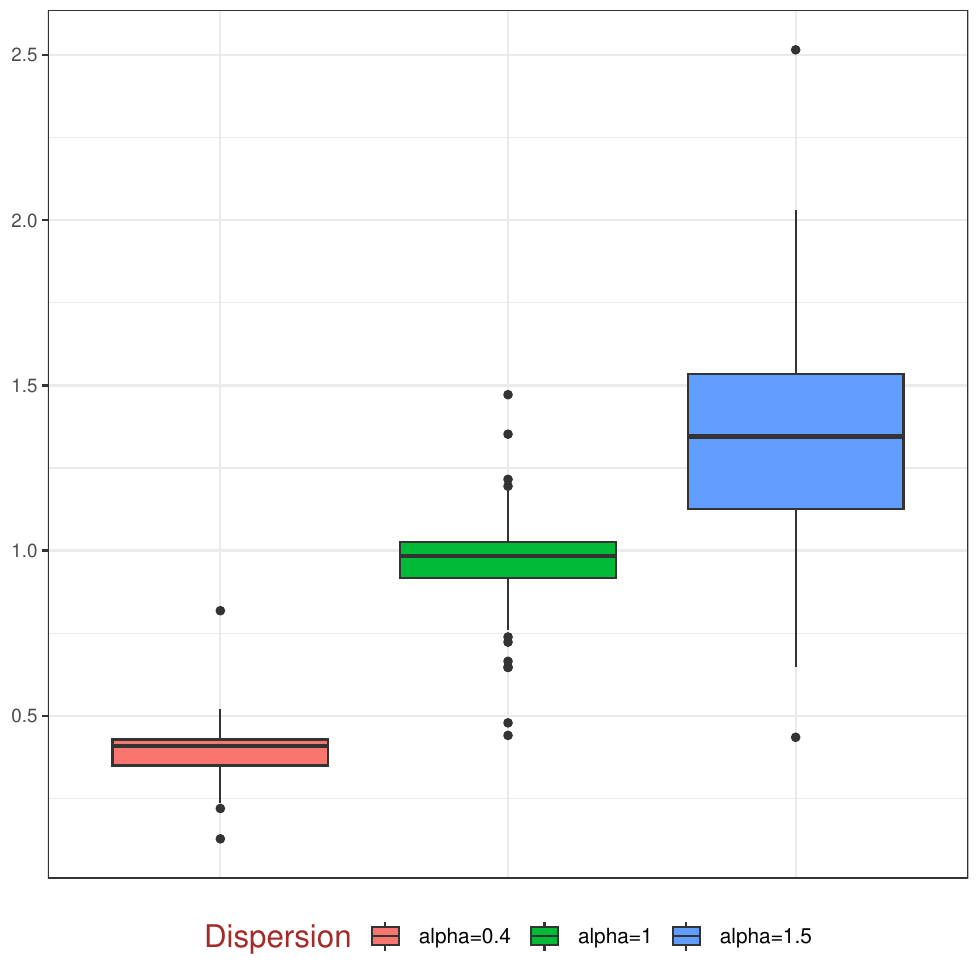}
    \caption{Simulation study: The boxplots of SE of $\hat{\alpha}$ for different models as well as dispersion scenarios (left) and the estimation of $\alpha$ (right).}
    \label{figmseA}
\end{figure}

\begin{figure}
    \centering
      \includegraphics[scale = 0.52]{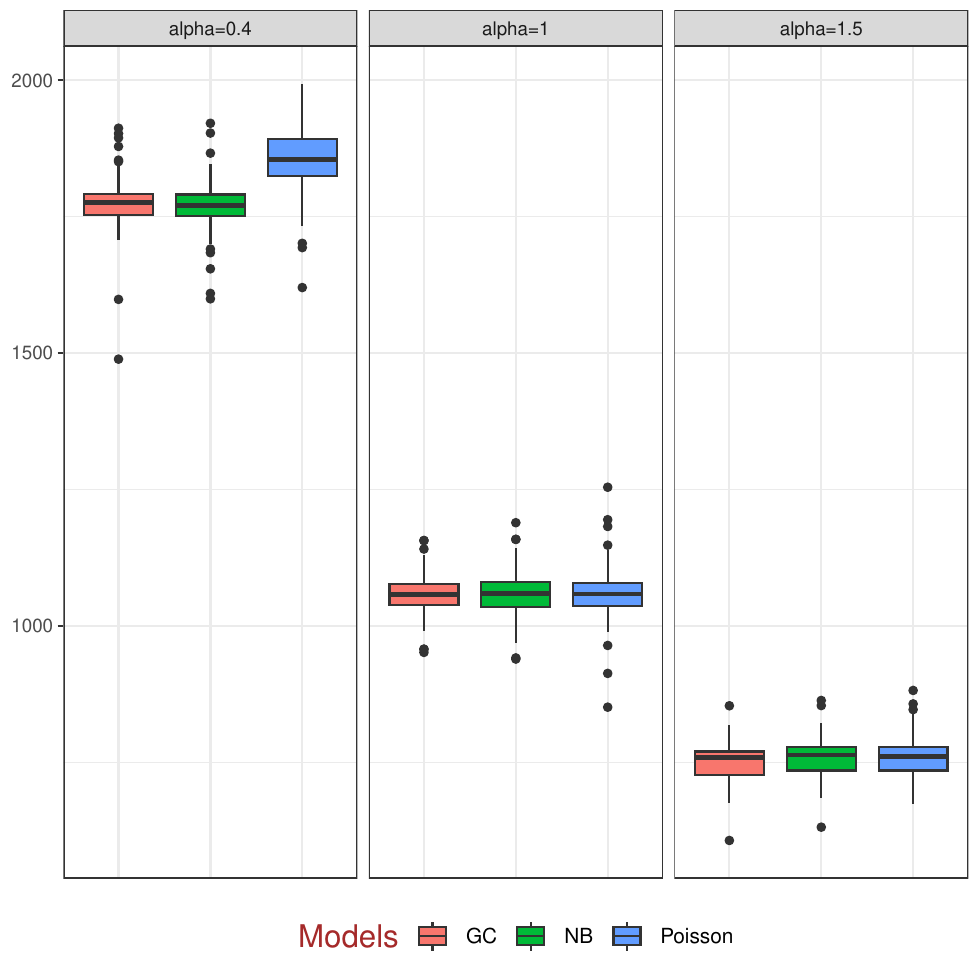} 
    \includegraphics[scale = 0.52]{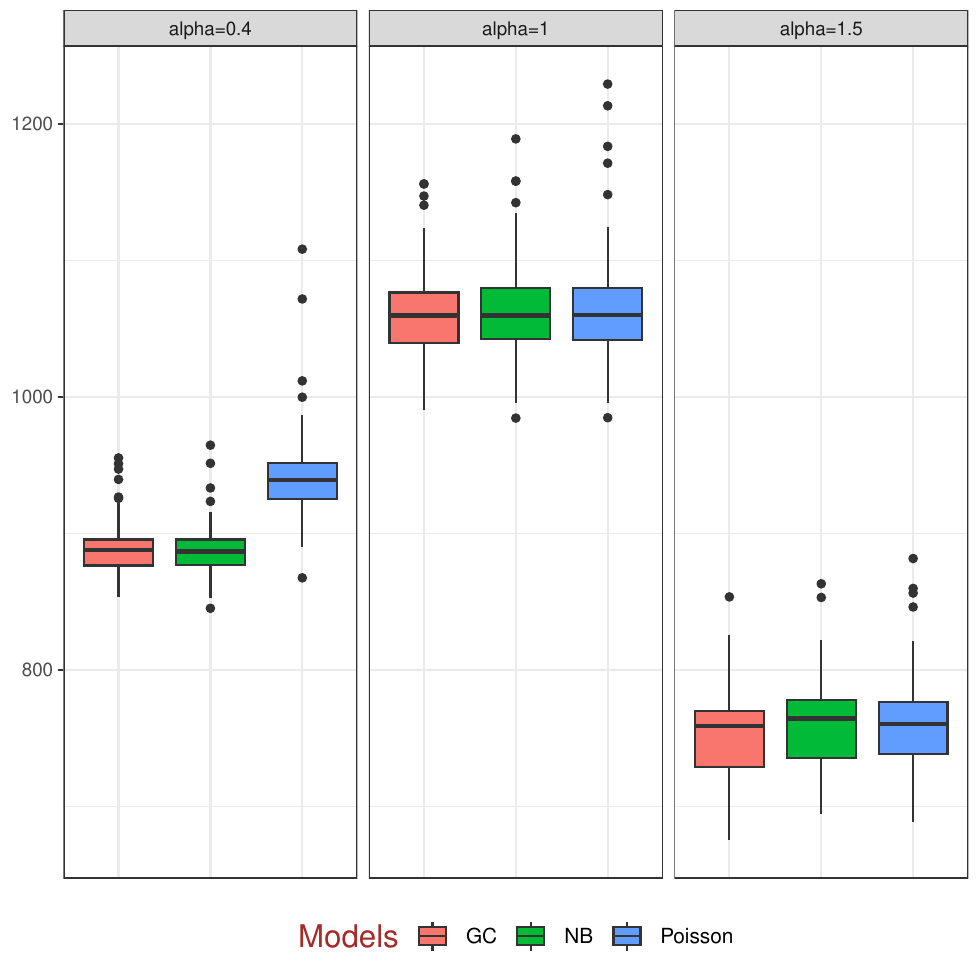}
    \caption{Simulation study: The boxplots of DIC (left) and WAIC (right) with $f(x)=\sin(x)$ for different models well as dispersion scenarios. \label{figdic}}
\end{figure}

The  MSEs for the univariate and bivariate functional effects are represented in Figures \ref{figmsefz} and \ref{figmsefs}, respectively. Corresponding SEs and estimates of $\hat{\alpha}$ are also given in Figure \ref{figmseA}.  Figure \ref{figdic} shows the values of DIC and WAIC. In conclusion, we can summarise the results as follows:

The MSE values in Figure \ref{figmsefz} exhibit minimal variation among the various count models in capturing fixed effects. This indicates that the models have similar performance. The GC model shows better effectiveness in preserving spatial relationships, especially in cases of over-dispersion, as seen in Figure \ref{figmsefs}. Although the NB model performs well with over-dispersed data, the GC model is more robust. When dispersion is equivalent, all models perform similarly. However, in scenarios of under-dispersion, the GC model demonstrates significantly better performance. Figure \ref{figmseA} illustrates that while the overall estimations of the dispersion parameter, $\alpha$, are satisfactory in all three scenarios, there is increased uncertainty in cases of under-dispersion due to larger variance.
In addition, the visual depiction of DIC and WAIC values in Figure \ref{figdic} effectively emphasizes the reliability of our proposed model in various scenarios, particularly when dealing with situations of over-dispersion and under-dispersion.

\section{Conclusion}\label{Sec6}
Count data often displays significant dispersion and intricate spatial dependencies that traditional models, including Poisson and NB regression, do not effectively address. The complex and poorly understood interactions between fixed effects and spatial random effects are frequently oversimplified, resulting in biased inferences and diminished predictive accuracy. We propose a semi-parametric spatial GC regression model that integrates renewal theory to accommodate non-exponential waiting times between events, while employing nonparametric strategies to manage complex spatial structures and varying dispersion levels. This method improves model adaptability by enabling the concurrent estimation of nonlinear fixed effects and spatial dependencies, while maintaining the fundamental count data characteristics of the GC process.

The proposed model's performance is assessed through three distinct case studies, each illustrating its capability to manage real-world spatial count data characterized by varying levels of dispersion and complexity. The initial example functions as a benchmark study, commonly utilized in nonparametric spatial modeling \citep{wood2017, yue2014}, offering a validation framework for the comparison of our method with established approaches. The second application examines mortality rates from lung and bronchus cancer across various spatial regions, integrating environmental covariates like air pollution and health factors. This analysis demonstrates that our model effectively captures spatial heterogeneity and nonlinear fixed effects, aspects that traditional count models inadequately address. The findings support the model's relevance in public health and epidemiological research, emphasizing the importance of comprehending spatial disease patterns. This example analyzes the frequency of days with a minimum of 1.0 mm of precipitation in Alberta, Canada, for May 2024. This application is significant in climate science, as predicting precipitation extremes is essential for comprehending the impacts of climate change and informing resource management strategies. The model's capacity to address missing data in spatial predictions underscores its utility in environmental research. 

Our model demonstrates superior performance compared to traditional Poisson, NB, and generalized Poisson models across all applications, highlighting the necessity of addressing non-equivalent dispersion and spatial dependencies in count data modeling. We conducted a simulation study to evaluate the model’s performance, confirming its capacity to accurately estimate dispersion, recover complex spatial patterns, and enhance predictive power compared to existing alternatives. Using semi-parametric spatial modeling along with generalized count properties creates a strong and adaptable framework for analyzing count data that has a lot of complex dependencies. The applicability is notably significant in disciplines like environmental science, epidemiology, and spatial statistics, where precise identification of nonlinear effects and spatial structures is crucial for dependable inference.

\section*{Acknowledgment}
This work was based upon research supported in part by the National Research Foundation (NRF) of South Africa (SA), NRF ref. SRUG2204203865 and ref. RA211204653274, grant No. 151035, as well as the Centre of Excellence in Mathematical and Statistical Sciences, based at the University of the Witwatersrand (SA). The opinions expressed and conclusions arrived at are those of the authors and are not necessarily to be attributed to the NRF. Mohammad Arashi's work is based on research supported in part by the Iran National Science Foundation (INSF) grant No. 4015320.

\newpage
\bibliographystyle{apalike}  

\end{document}